\documentclass[12pt]{svjour2}

\smartqed
\usepackage{graphics,graphicx,subfigure,adjustbox,amssymb,xcolor}
\usepackage{amsmath}

\journalname{J Stat Phys}

\begin{document}

\title{Low-temperature anomalies in one-dimensional exactly solvable fluids}  

\titlerunning{Low-temperature anomalies in one-dimensional fluids}

\author{Igor Trav\v{e}nec $\bullet$ Ladislav \v{S}amaj}

\institute{Institute of Physics, Slovak Academy of Sciences, 
D\'ubravsk\'a cesta 9, SK-84511 Bratislava, Slovakia \\
\email{Igor.Travenec@savba.sk and Ladislav.Samaj@savba.sk}}

\date{Received:  / Accepted: }

\maketitle

\begin{abstract}
Previous experiments and numerical simulations have revealed that a
limited number of two- and three-dimensional particle systems contract
in volume upon heating isobarically.
This anomalous phenomenon is known as negative thermal expansion (NTE).
The present paper focuses on the possibility of NTE in exactly solvable
one-dimensional fluids.
Firstly, the quantization of classical pure hard rods (free of NTE)
does not induce NTE which indicates an unimportant role of quantum mechanics
in the topic.
Secondly, the classical hard rods with various types of soft nearest-neighbor
interactions that contain a basin of attraction with only one minimum
are investigated.
The ground-state analysis reveals that, for certain potentials, increasing
the pressure can lead to a discontinuous jump in the mean spacing between
particles.  
The low-temperature analysis of the exact equation of state indicates
that the NTE anomaly is present if the curvature of the soft
potential increases with the distance between particles or if
the potential exhibits a singularity within the basin of attraction.
Isotherms of the compressibility factor, which measures the deviation of
the thermodynamic behavior of a real gas from that of an ideal gas,
demonstrate typical plateau or double-plateau shapes in large intervals of
particle density.


\end{abstract}

\renewcommand{\theequation}{1.\arabic{equation}}
\setcounter{equation}{0}

\section{Introduction} \label{Sec1}
Standard solid and fluid materials expand in volume when heated under
constant pressure.
However, there are special materials that contract in volume upon heating;
this anomalous phenomenon is known as negative thermal expansion (NTE).

NTE has been experimentally observed in three-dimensional (3D) substances with
anisotropic intermolecular pairwise interactions.
One well-known example is ice under atmospheric pressure which exhibits
NTE at low temperatures $\lesssim 63 K$ \cite{Rottger94,Evans99}
as well as upon melting ice into liquid water in the temperature interval
$0-3.98$ $^\circ C$ \cite{Greenwood97,Errington01}.
NTE has also been detected in graphene \cite{Yoon11} and
complex compounds like zirconium tungstate Zr${\rm W}_2{\rm O}_8$
\cite{Mary96}, diamond and zinc-blende semiconductors \cite{Biernacki89},
${\rm Lu}_2{\rm W}_3{\rm O}_{12}$ \cite{Forster98}, etc.

NTE has been observed in computer simulations of two-dimensional (2D)
and 3D single-component classical (i.e., non-quantum) systems with
{\em isotropic} pair interactions \cite{Rechtsman07} provided that
the soft-core potential $\varphi(r)$ has a region of attraction with
increasing curvature.
In one dimension (1D), the necessary and sufficient condition for NTE 
as stated by Kuzkin \cite{Kuzkin14} is
\begin{equation} \label{Kuzkin}
\varphi'''(l)>0 ,  
\end{equation}
where $l$ is the mean spacing between particles at a given temperature.

Other models of identical particles with {\em repulsive}
isotropic interactions that exhibit NTE also exist.
NTE arises at low temperatures in 2D systems with a specific topography of
the energy landscape \cite{Batten09a,Batten09b}.
The NTE anomaly was detected using computer simulations in 3D
repulsive potentials with two competing length scales:
the hard-core diameter $a$ and a finite soft-core range $a'>a$,
see references \cite{Sadr98,Velasco00,Ryzhov03,Kumar05,Xu06,Yan06,Oliveira08,Oliveira09,Buldyrev09,Zhou09,Coquand20}.
The repulsive soft-core potentials used in simulations were mainly
the square shoulder potential \cite{Hemmer70,Stell72} and the linear
ramp potential of Jagla \cite{Jagla99}, or smoothed
potentials \cite{Sadr98,Oliveira08,Oliveira09,Fomin08,Gribova09}.
It is generally believed that repulsive potentials with two length scales
always exhibit water-like anomalies \cite{Oliveira09,Fomin08}.

NTE occurs in the 3D Gaussian core model with an interaction potential
that is analytic everywhere \cite{Stillinger97}.

In one dimension (1D), there is a large family of classical fluids
consisting of particles that interact pairwise through a potential
which are exactly solvable in thermal equilibrium at temperature $T$.

The simplest 1D system of this type is the Tonks gas of hard rods with
the interaction potential
\begin{equation} \label{hr}
\phi(x) = \left\{
\begin{array}{ll}
\infty & \mbox{if $\vert x\vert < a$,} \cr
0 & \mbox{if $\vert x\vert \ge a$,}  
\end{array} \right.   
\end{equation}
where $a$ represents the diameter of the hard core around each particle.
For the classical version of this model, Tonks \cite{Tonks36} derived
the equation of state (EoS) which relates the particle density to
temperature $T$ and pressure $p\ge 0$.
This simple model is free of NTE anomaly.
The two-body distribution function of classical hard rods was calculated
earlier \cite{Zernike27}.
The quantum version of 1D hard rods was solved exactly by using the Bethe
ansatz in reference \cite{Wadati02}, see also textbook \cite{Samaj13}.

Takahashi \cite{Takahashi42} solved a more general classical model with
interactions only among nearest neighbors using the canonical ensemble.
Bishop and Boonstra \cite{Bishop83} performed a rederivation of the EoS
for the Takahashi gas within the isothermal-isobaric ensemble. 
The many-particle distribution functions have been
derived in reference \cite{Salsburg53}.

Interactions among all particles were reduced in the Takahashi gas
to the nearest-neighbor pairs approximately ad-hoc ``by hand''.
Let us now consider 1D hard rods of diameter $a$ interacting
via the two-length-scales pair potential
\begin{equation} \label{general}
\phi(x) = \left\{
\begin{array}{ll}
\infty & \mbox{if $\vert x\vert < a$,} \cr
\varphi(x) & \mbox{if $a\le \vert x\vert < a'$,} \cr
0 & \mbox{if $\vert x\vert \ge a'$,}  
\end{array} \right.   
\end{equation}  
where $\varphi(x)=\varphi(-x)$ and the range $a'$ of the soft-core potential
$\varphi(x)$ is restricted by $a'\le 2a$.
For a given particle, this restriction implies that hard cores of
first neighbors prevent interaction with second neighbors,
effectively reducing of the interaction potential to nearest
neighbors only without any ad-hoc assumptions.
The exact solution of thermal equilibrium of 1D fluids with nearest-neighbor
interactions (\ref{general}) includes systems of identical particles together
with mixtures \cite{Lebowitz71,Heying04,Santos07,Ben-Naim09,Sahnoun24},
see books \cite{Percus87,Santos16}.
A scaling regime in which the system exhibits a second-order phase transition
was proposed by Jones \cite{Jones85}.

Recently\cite{Travenec25}, exactly solvable 1D fluids of hard rods with
various types of soft {\em purely repulsive} nearest-neighbor interaction
potentials $\varphi(x)>0$, such as the square shoulder, linear
and quadratic ramps and certain logarithmic potentials
were examined for the NTE phenomenon in the low-temperature region.
These particle systems were found to exhibit NTE anomalies, with
the presence of the anomaly depending greatly on the shape of
the core-softened potential.
In some cases, NTE was associated with equidistant ground states jumps in
chain spacing $a'\to a$ at specific pressures.
Kuzkin's necessary and sufficient condition for NTE (\ref{Kuzkin}) 
did not apply to these systems.
Isotherms of the compressibility factor, which measures the deviation of the
thermodynamic behavior of a real gas from an ideal gas, were shown to display
both monotonous and non-monotonous behaviors as functions of the particle
density, with a paradoxical weakening of pressure as density increases.

\begin{figure}[t]
\begin{center}
\includegraphics[clip,width=0.84\textwidth]{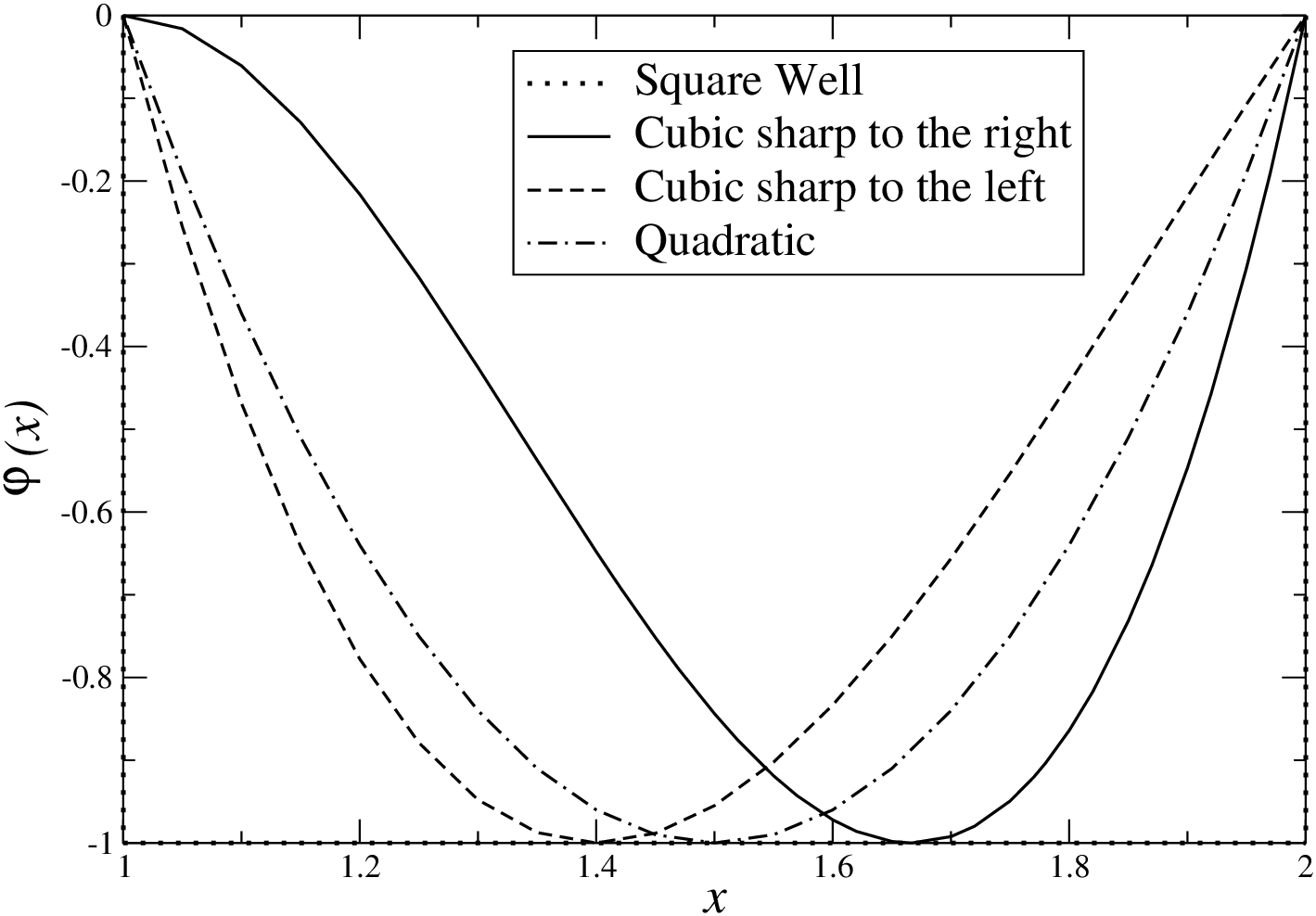}
\caption{Some of the interaction potentials considered in this work:
the square well potential and polynomial models.
The length scale with the hard-core diameter $a=1$ is used.}
\label{fig1}
\end{center}
\end{figure}

The present paper continues in focusing on the possibility of NTE
in exactly solvable 1D fluids.
Firstly, we study the classical and quantum versions of pure hard rods
and show that the inclusion of quantum fluctuations in classical
model (which itself is free of NTE) does not induce NTE.
This indicates an unimportant role of quantum mechanics in the topic.
Secondly, classical hard rods with soft nearest-neighbor
interaction potentials that contain a basin of attraction with only
one minimum (except for the square well potential) are considered.
Some of the soft potentials are depicted in figure \ref{fig1}.
The ground-state and low-temperature analysis of EoS indicates that NTE
anomaly is present if the curvature of the interaction
potential increases with the distance between particles, in agreement
with Kuzkin's condition (\ref{Kuzkin}).
Another possible cause for NTE is a singularity of the potential
within the basin of attraction.
Isotherms of the compressibility factor show to exhibit plateau or
double-plateau shapes in large intervals of particle density.

The paper is organized as follows.
Section \ref{Sec2} discusses the general formalism for exactly solvable
1D fluids with nearest-neighbor interactions of the type (\ref{general})
with $a'\le 2a$, in the ground state at $T=0$ or at arbitrary $T$.
The compressibility factor is introduced. 
The derivation and analysis of EoS for the classical and
quantum versions of Tonks model (\ref{hr}) are presented
in section \ref{Sec3}.
The square well model with a constant negative potential between
$a$ and $a'$ is studied in section \ref{Sec4}.
The polynomial (quadratic and cubic) models
are analyzed in section \ref{Sec5}.
Section \ref{Sec6} deals with interaction potentials that exhibit
a singularity inside the basin of attraction, specifically at, to the left
and to the right of the minimum point. 
Section \ref{Sec7} provides a brief recapitulation and concluding remarks.

\renewcommand{\theequation}{2.\arabic{equation}}
\setcounter{equation}{0}

\section{General formalism for 1D classical fluids} \label{Sec2}

\subsection{Exact formulas for the EoS} \label{Sec2.1}
$N$ identical particles of the 1D classical fluid move in a continuous box
of length $L$, say with periodic boundary conditions.
The thermodynamic limit $L\to\infty$ and $N\to\infty$, while keeping
the particle number density $n=N/L$ fixed, is considered.
The pair potential $\phi(x)$ is of the form (\ref{general}) with
the restriction $a'\le 2 a$, ensuring that each particle interacts
only with its nearest neighbors.
The hard-core condition $\phi(x)=\infty$ for $\vert x\vert \in [0,a]$
fixes the order of the particles on the line.

The soft-core potential $\varphi(x)$ is supposed to be attractive,
\begin{equation}
\varphi(x) \le 0 , \qquad x\in [a,a'] ,
\end{equation}  
with just one minimum at point $x=a_m$ $(a<a_m<a')$ at which its derivative
with respect to $x$ vanishes:
\begin{equation}
\frac{\partial\varphi(x)}{\partial x} \Big\vert_{x=a_m} = 0 .
\end{equation}  
The only exception from this condition is the square well potential
(section \ref{Sec4}) possessing an infinitely degenerate minimum.
The derivative is assumed to be negative for $x\in (a,a_m)$ and
positive for $x\in (a_m,a')$:
\begin{equation}
\frac{\partial\varphi(x)}{\partial x} < 0 \quad
\mbox{for $x\in (a,a_m)$}, \qquad
\frac{\partial\varphi(x)}{\partial x} > 0 \quad \mbox{for $x\in (a_m,a')$}.
\end{equation}  
Moreover, the potential is assumed to be continuous at the interaction
border $x=a'$,
\begin{equation}
\varphi(a') = 0.
\end{equation}
Without any loss of generality, in numerical evaluations of exact
formulas the soft-core potential is taken to vanish also in the hard-core
limit as $x\to a^+$,
\begin{equation}
\varphi(x\to a^+) = 0.
\end{equation}
The value of the potential at the minimum point $\varphi(a_m)$ will
usually be fixed to $-1$.

Particles are in thermal equilibrium with the thermostat at temperature $T$, 
or inverse temperature $\beta=1/(k_{\rm B}T)$ where the Boltzmann constant
$k_{\rm B}$ will be set to unity.
The model is exactly solvable in the isothermal-isobaric ensemble,
see reference \cite{Santos16} for the notation.
Instead of the particle density $n$, EoS will be formulated in terms of
the averaged distance between particle (reciprocal density) $l=1/n$.
The exact EoS can be expressed in terms of the Laplace transform of
the pair Boltzmann factor ${\rm e}^{-\beta\phi(x)}$,
\begin{equation}\label{Om}
\widehat{\Omega}(s) = \int_0^{\infty} {\rm d}x\,
{\rm e}^{-xs} {\rm e}^{-\beta\phi(x)} 
\end{equation}
and of its derivative
\begin{equation}
\widehat{\Omega}'(s) \equiv \frac{\partial\widehat{\Omega}(s)}{\partial s}
= - \int_0^{\infty} {\rm d}x\, x {\rm e}^{-xs} {\rm e}^{-\beta\phi(x)}
\end{equation}  
as follows:
\begin{equation}\label{recdensity}
l(\beta,p) = - \frac{\widehat{\Omega}'(\beta p)}{\widehat{\Omega}(\beta p)}
= \frac{\int_0^{\infty} {\rm d}x\, x {\rm e}^{-\beta p x} {\rm e}^{-\beta\phi(x)}}{
\int_0^{\infty} {\rm d}x\, {\rm e}^{-\beta p x} {\rm e}^{-\beta\phi(x)}} . 
\end{equation}  
For the two-length-scales pair potential (\ref{general}), this
formula can be written as
\begin{equation} \label{crushformula}
l(\beta,p) = - \frac{1}{\beta} \frac{\partial}{\partial p}
\ln \left[ \int_a^{a'} {\rm d}x\, {\rm e}^{-\beta f(x)} +
\frac{{\rm e}^{-\beta p a'}}{\beta p} \right] ,  
\end{equation}
where
\begin{equation} \label{f}
f(x) \equiv \varphi(x) + p x  
\end{equation}
and $p>0$; the dependence of $f(x)$ on $p$ will not be explicitly expressed
to save space.

The integral in the exact formula for the mean spacing (\ref{crushformula}) 
can be expressed in terms of elementary or special functions for
certain forms of the soft-core potential $\varphi(x)$.
In that case the ground-state and low-$T$ (large-$\beta$) analyses can be
easily done explicitly.
In the opposite case, the ground-state analysis and a systematic low-$T$
expansion can be done for general $\varphi(x)$ by using a procedure
outlined below.

\subsection{Ground state and the leading low-$T$ correction} \label{Sec2.2}
In general, it is a difficult task to prove that the ground state of
interacting classical particles is periodic \cite{Radin87,Bris05,Blanc15}.
The situation is simpler in 1D fluids where a number of theorems have been
proven for various kinds of interaction
potentials \cite{Ventevogel78,Ventevogel79a,Ventevogel79b}.
For the considered hard-core potential with a finite-range soft
potential (\ref{general}), Radin and Schulman proved the periodicity
of the 1D ground state in reference \cite{Radin83}.

The ground-state formula for the mean spacing $l$ and its low-$T$
(large-$\beta$) corrections can be systematically generated from
the formula (\ref{crushformula}) by using the standard saddle-point method
applied to the function $f(x)$ in the integral under the logarithm.
Let the function $f(x)$ have its minimum at $x^*$, i.e.,
\begin{equation}
\frac{\partial f(x)}{\partial x}\Big\vert_{x=x^*} = 0 , \qquad
\frac{\partial^2 f(x)}{\partial x^2}\Big\vert_{x=x^*} > 0 .  
\end{equation}
In view of definition (\ref{f}), this condition is equivalent to the one
\begin{equation} \label{derener}
p = - \frac{\partial\varphi(x)}{\partial x}\Big\vert_{x=x^*} .
\end{equation}
The inversion of this relation defines the ground-state function $x^*(p)$.
Inserting the leading term of the Taylor expansion
\begin{equation}
f(x) = f(x^*) + O\left( \left( x-x^* \right)^2 \right)  
\end{equation}
into the integral in (\ref{crushformula}), one gets
\begin{equation} \label{crucequation}
l(\beta,p) \approx \frac{\partial}{\partial p} f\left( x^*(p)\right)
- \frac{1}{\beta} \frac{\partial}{\partial p} \ln \left\{ (a'-a)
+ \frac{1}{\beta p} {\rm e}^{-\beta\left[ p(a'-x^*)-\varphi(x^*)\right]} \right\} . 
\end{equation}
The second term on the rhs of this equation is exponentially small,
specifically of type $\exp(-c\beta)$ with $c>0$, and therefore can be neglected
in the limit $\beta\to\infty$. 
The first term on the rhs of (\ref{crucequation}) can be written as
\begin{eqnarray}
\frac{\partial}{\partial p} f\left( x^*(p)\right) & = &
\frac{\partial}{\partial p} \left[ \varphi\left( x^*(p)\right) +
p x^*(p) \right] \nonumber \\
& = & \frac{\partial\varphi(x^*)}{\partial x^*} \frac{\partial x^*}{\partial p}
+ x^*(p) + p \frac{\partial x^*}{\partial p} \nonumber \\
& = & x^*(p) ,
\end{eqnarray}  
where the transition from the second to the third line follows from
(\ref{derener}).
It is clear that $x^*(p)$ determined by equation (\ref{derener}) is nothing
but the ground-state spacing,
\begin{equation} \label{gsspacing}
l(T=0,p) = x^*(p).
\end{equation}  
In the limit $p\to 0^+$, $x^*$ coincides with the minimum point of
the potential $\varphi(x)$,
\begin{equation}
\lim_{p\to 0^+} l(T=0,p) = a_m .
\end{equation}
On the other hand, $x^*(p)$ attains its hard-core minimum equal to $a$
at the ``incompressibility'' pressure $p_i$ given by
\begin{equation} \label{pi}
p_i = - \frac{\partial\varphi(x)}{\partial x}\Big\vert_{x=a} .
\end{equation}
Due to the presence of the hard core, the ground-state spacing remains
to be equal to $a$ also for all pressures larger than $p_i$,
\begin{equation}
l(T=0,p) = a \qquad \mbox{for all $p\ge p_i$.}
\end{equation}

To obtain the leading low-$T$ correction to the ground-state formula
(\ref{gsspacing}) one must consider the next term of
the Taylor expansion of $f(x)$ around the point $x^*$,
\begin{equation}
f(x) = f(x^*) + \frac{1}{2!} \varphi''(x^*) \left( x-x^* \right)^2 +
O\left( \left( x-x^* \right)^3 \right) ,  
\end{equation}
where the equality $f''(x)=\varphi''(x)$ was utilized.
Neglecting terms of order $\left( x-x^* \right)^3$ and substituting
the above expansion into the integral in (\ref{crushformula}), one obtains:
\begin{equation} \label{cruceq}
l(\beta,p) \approx  x^*(p) - \frac{1}{\beta} \frac{\partial}{\partial p}
\ln \Bigg\{ \int_a^{a'} {\rm d}x\, {\rm e}^{-\frac{\beta}{2}\varphi''(x^*)(x-x^*)^2}
+ \frac{1}{\beta p} {\rm e}^{-\beta\left[ p(a'-x^*)-\varphi(x^*)\right]} \Bigg\} . 
\end{equation}
By using the substitution $y=\sqrt{\beta\varphi''(x^*)/2}(x-x^*)$,
the integral in the second term on the rhs of (\ref{cruceq})
can be transformed to
\begin{equation}
\sqrt{\frac{2}{\beta\varphi''\left( x^*(p)\right)}}  
\int_{\sqrt{\beta\varphi''(x^*)/2}(a-x^*)}^{\sqrt{\beta\varphi''(x^*)/2}(a'-x^*)}
{\rm d}y\, {\rm e}^{-y^2} .
\end{equation}
In view of the inequalities $(a'-x^*)>0$ and $(a-x^*)<0$, in the considered
$\beta\to\infty$ limit the lower and upper bounds of integration can be
substituted by $-\infty$ and $\infty$, respectively.
Since $p(a'-x^*)-\varphi(x^*)>0$ for the attractive soft-core potential
$\varphi(x)\le 0$, the integral is dominant with respect to the exponentially
small term under the logarithm in equation (\ref{cruceq}), resulting in
the low-$T$ expansion:
\begin{equation} \label{lpT}
l(T,p) = x^*(p) + \frac{T}{2}\frac{\partial}{\partial p}
\ln\left({\frac{\partial^2 \varphi(x)}{\partial x^2}\Big\vert_{x=x^*(p)}}\right)
+ o(T) .
\end{equation}

The validity of the analysis above is limited to pressures $0\le p<p_i$.
The reason is that for $p\ge p_i$ the coordinate $x=a$ no longer yields
the minimum of $f(x)\equiv\varphi(x)+px$.
Therefore, the Taylor expansion of $f(x)$ around the point $x=a$
also includes the linear term:
\begin{equation} \label{Taylorexp}
f(x) = f(a) + (p-p_i)(x-a) + \frac{1}{2} \varphi''(a) (x-a)^2 + \cdots ;   
\end{equation}
here, we have utilized the definition of $p_i$ (\ref{pi}) and
the evident equality $f''(x)=\varphi''(x)$.
By substituting the Taylor expansion (\ref{Taylorexp}) into the integral
in (\ref{crushformula}) and employing the substitution $y=x-a$,
we obtain
\begin{equation} \label{crucequat}
l(\beta,p) \approx a - \frac{1}{\beta} \frac{\partial}{\partial p}
\ln \left\{ \int_0^{a'-a} {\rm d}y\,
{\rm e}^{-\beta(p-p_i)y-\frac{\beta}{2}\varphi''(a)y^2}
+ \frac{1}{\beta p} {\rm e}^{-\beta\left[ p(a'-a)-\varphi(a)\right]} \right\} . 
\end{equation}
Since $p(a'-a)-\varphi(a)>0$, the exponentially small term can be neglected
with respect to the integral in the limit $\beta\to\infty$ and one arrives at 
\begin{equation} \label{ratio} 
l(\beta,p) \approx a + \frac{\int_0^{a'-a} {\rm d}y\, y
{\rm e}^{-\beta(p-p_i)y-\frac{\beta}{2}\varphi''(a)y^2} }{
\int_0^{a'-a} {\rm d}y\, {\rm e}^{-\beta(p-p_i)y-\frac{\beta}{2}\varphi''(a)y^2}} .  
\end{equation}  
\begin{itemize}
\item
For $p=p_i$, the term linear in $y$ disappears and one finds in
the $\beta\to\infty$ limit that
\begin{equation} \label{lpiT}
l(T,p_i) = a + \sqrt{\frac{2}{\pi\varphi''(a)}} \sqrt{T} + o(\sqrt{T}) . 
\end{equation}
Note the transition from the analytic expansion in $T$ for $0\le p<p_i$
(\ref{lpT}) to a singular expansion in $\sqrt{T}$ at $p=p_i$.
\item
When $p>p_i$, the terms linear in $y$ dominate over the quadratic ones
and one gets
\begin{equation} \label{lpiTgreater}
l(T,p) = a + \frac{1}{p-p_i} T + o(T) , \qquad p>p_i . 
\end{equation}
The prefactor to $T$ diverges as $p$ goes to $p_i$ from above which is
a sign of the change in the analytic form of the leading
low-$T$ correction from $T$ for $p>p_i$ to $\sqrt{T}$ at $p=p_i$.
Notice that an infinite pressure has to be applied to squeeze the particles
to their smallest hard-core distance $a$ if $T>0$, in contrast to the
$T=0$ ground state with $l(0,p)=a$ for all $p\ge p_i$.
\end{itemize}

The scenario above holds if $x^*(p)$ changes continuously from $a_m$ at
$p\to 0$ to $a$ at $p=p_i$.
Possible exceptions associated with a discontinuity of $x^*(p)$ will
be discussed in particular cases in section \ref{Sec5}.

\subsection{The compressibility factor} \label{Sec2.3}
The deviation of thermodynamic behavior of real gases from the EoS
of an ideal gas $\beta p = n$ is measured by the compressibility factor $Z$
defined as:
\begin{equation} \label{Z}
Z(\beta,n) \equiv \frac{\beta p}{n} .
\end{equation}
In the low-density limit as $n\to 0$, it holds that
\begin{equation}
Z(\beta,n) \mathop{\sim}_{n\to 0} 1 .
\end{equation}
Another regime that can be treated rigorously is the neighborhood
of the close packed limit $l\to a^+$ (or $n a\to 1^-$) when
$\beta p\to \infty$ \cite{Travenec25}.
In particular, 
\begin{eqnarray}
Z(\beta,n) & = & \frac{1}{1-an} -\beta \varphi'(a) a
-\beta \left[ \varphi'(a) a + 2 \varphi''(a) a^2 \right] (1-an)
\nonumber \\ & & + O\left[(1-an)^2\right] . \label{Znan} 
\end{eqnarray}
The leading singular term of this expansion is universal, i.e.,
independent of the model's parameters, except for the hard core diameter $a$.
The only known exception from this singular behavior is the sticky balls
model \cite{Percus87,Santos16} with the leading term of the form
$Z\propto 1/\sqrt{1-a n}$.

\renewcommand{\theequation}{3.\arabic{equation}}
\setcounter{equation}{0}

\section{1D Tonks gas} \label{Sec3}

\subsection{Classical hard rods} \label{Sec3.1}
For the classical 1D fluid of hard rods with the interaction potential
(\ref{hr}), the Laplace transform of the pair Boltzmann factor (\ref{Om})
reads as
\begin{equation}
\widehat{\Omega}(s) = \frac{{\rm e}^{-a s}}{s} .
\end{equation}
By applying (\ref{recdensity}), the EoS for the reciprocal density takes
the form
\begin{equation} \label{EoShr}
l(T,p) = a + \frac{T}{p}.
\end{equation}
When $a$ approaches zero (pointlike particles), the EoS of the ideal gas
$\beta p = n$ is obtained.

In the low-temperature limit $T\to 0$ and for any positive pressure $p>0$,
the particle spacing $l(0,p)$ becomes the lattice constant $a$ of
the close packed array of hard rods in the ground state.
As temperature $T$ increases, the average distance between particles
grows linearly with $T$ and diverges in the high-temperature limit
$T\to\infty$.
This behavior is due to thermal fluctuations which weaken energy bounds
among particles.
NTE anomaly is absent for classical 1D Tonks gas.

The compressibility factor is given by
\begin{equation}
Z = \frac{\beta p}{n} = \frac{1}{1-a n} .
\end{equation}
It involves only the leading term of the expansion (\ref{Znan}). 

\subsection{Quantum hard rods} \label{Sec3.2}
The exact solution of 1D quantum hard rods \cite{Wadati02} was given as
the simplest example of general formalism of analytic and thermodynamic
Bethe ansatz in worked-out exercises of textbook \cite{Samaj13}.
Hard rods are taken as identical spinless particles of mass $m$,
it does not matter whether bosons or fermions (the relationship between
the particle spin and Fermi/Bose statistics is purposely ignored in the field
of integrable systems).
In units of reduced Planck constant $\hbar=1$ and $2m=1$,
the thermal de Broglie wavelength reads as
\begin{equation} \label{Broglie}
\lambda_B \equiv \hbar \sqrt{\frac{2\pi\beta}{m}} = 2 \sqrt{\pi\beta} . 
\end{equation}
  
The analytic Bethe ansatz for 1D quantum hard rods is discussed in
section 1.6 of \cite{Samaj13}.

The zero-temperature EoS
\begin{equation} \label{quantumgs}
l(0,p) - a = \left( \frac{2\pi^2}{3p} \right)^{1/3}
\end{equation}  
is derived in exercise 2.4 of section 2 in textbook \cite{Samaj13}.
It is seen that the particle spacing in the quantum case depends on $p$.
Its classical hard-core value $l(0,p)=a$ is obtained only in the limit
$p\to\infty$ while $l(0,p)$ diverges as $p\to 0$.
These properties of $l(0,p)$ are related to the uncertainty principle
in quantum mechanics.

The finite-temperature EoS is derived within the grand canonical ensemble
in exercise 3.3 of section 3 in textbook \cite{Samaj13}.
Introducing the shifted chemical potential $\tilde{\mu} \equiv \mu - a p$
and the rescaled particle density
\begin{equation}
\tilde{n} \equiv \frac{n}{1-a n} = \frac{1}{l-a} ,
\end{equation}  
it holds that
\begin{eqnarray}
p & = & \frac{1}{\pi} \int_{-\infty}^{\infty} {\rm d}k\,
\frac{k^2}{{\rm e}^{\beta(k^2-\tilde{\mu})}+1} =  - \frac{1}{2\sqrt{\pi}\beta^{3/2}}
{\rm Li}_{3/2}\left( -{\rm e}^{\beta\tilde{\mu}} \right) \label{eq1} \\ 
\tilde{n} & = & \frac{1}{2\pi} \int_{-\infty}^{\infty} {\rm d}k\,
\frac{1}{{\rm e}^{\beta(k^2-\tilde{\mu})}+1} = - \frac{1}{2\sqrt{\pi}\beta^{1/2}}
{\rm Li}_{1/2}\left( -{\rm e}^{\beta\tilde{\mu}} \right) . \label{eq2}
\end{eqnarray}  
Here, the polylogarithm function ${\rm Li}_n(z)$ is defined in
the complex $z$-plane over the open unit disk as the series
\begin{equation}
{\rm Li}_n(z) = \sum_{k=1}^{\infty} \frac{z^k}{k^n} .
\end{equation}  
The relation between the deviation of the particle spacing $l-a=1/\tilde{n}$
and temperature $T=1/\beta$ at fixed pressure $p$ can be obtained via
calculation of the parameter $\tilde{\mu}$ from relation (\ref{eq1})
and its subsequent substitution into (\ref{eq2}).

The high-temperature and low-temperature virial expansions of the pressure $p$ 
are constructed in exercise 3.5 of section 3 in textbook \cite{Samaj13}.
At high temperatures one gets
\begin{equation} \label{ht}
\frac{\beta p}{\tilde{n}} = 1 + \frac{1}{2^{3/2}}
\left(\lambda_B \tilde{n} \right)
+ \left( \frac{1}{2} - \frac{2}{3^{3/2}} \right)
\left(\lambda_B \tilde{n} \right)^2 + O(1/T^{3/2}) .
\end{equation}
Note that since $\lambda_B\to 0$ in the high-temperature limit $\beta\to 0$,
the expansion (\ref{ht}) passes to the classical EoS (\ref{EoShr})
as it should be by the correspondence principle in quantum mechanics.
At low temperatures one gets
\begin{equation} \label{lt}
p = \frac{2\pi^2}{3} \tilde{n}^3 + \frac{1}{6\tilde{n}} T^2
+ \frac{1}{30 \pi^2 \tilde{n}^5} T^4 + O(T^6) .  
\end{equation}

\begin{figure}[t]
\begin{center}
\includegraphics[clip,width=0.84\textwidth]{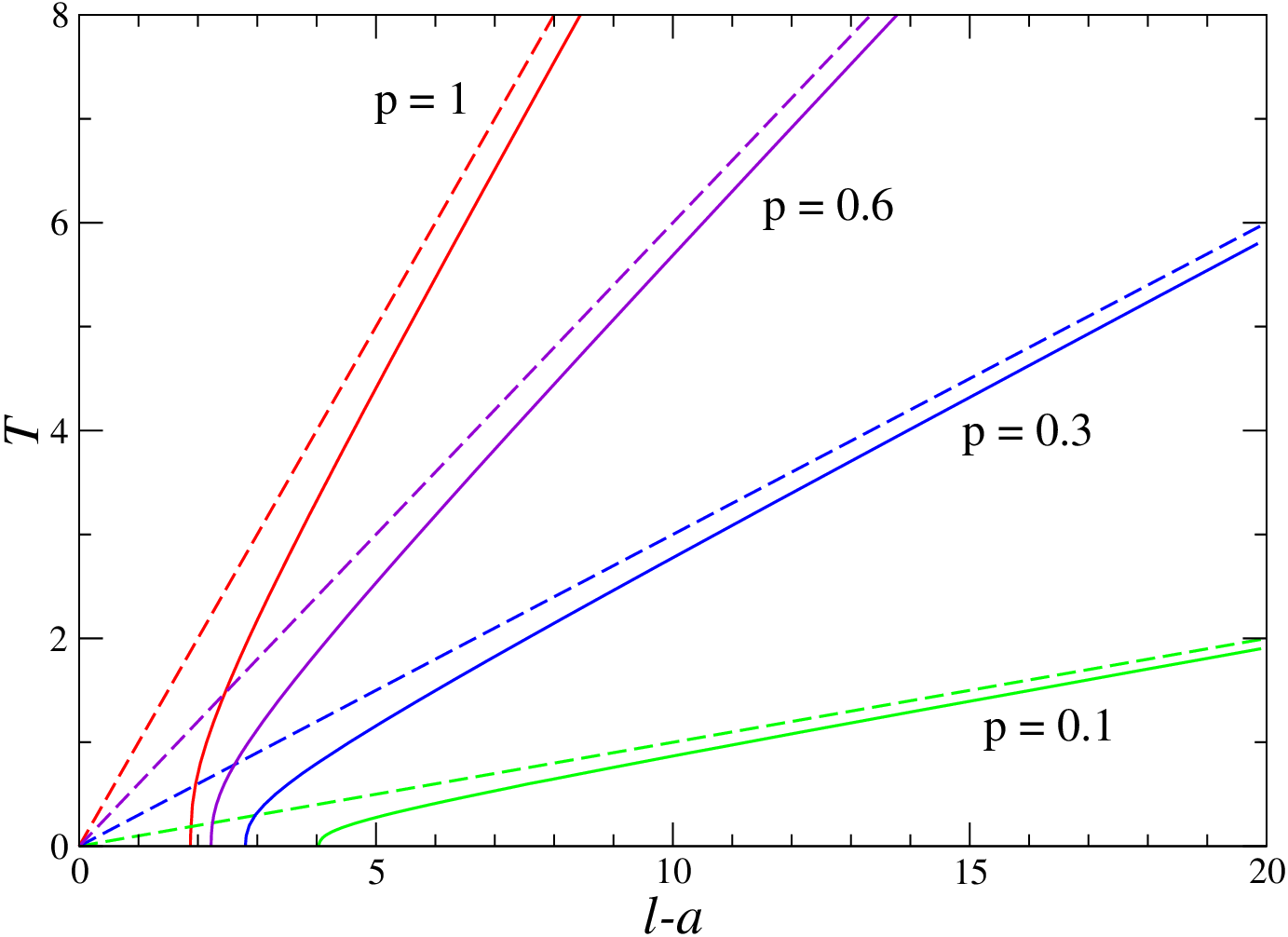}
\caption{1D system of pure hard rods.
The plots of temperature $T$ as the function of the shifted
mean spacing $l-a$ for the pressures $p=0.1, 0.3, 0.6$ and $1$.
The solid/dashed curves correspond to the quantum/classical version
of the model.}
\label{fig2}
\end{center}
\end{figure}

In this paper, instead of presenting the plots of $l$ versus $T$ for fixed $p$
we prefer to present the plots $T$ versus $l$ (or $l-a$) for fixed $p$ because
the NTE phenomenon is better seen in this format.
The plots of $T$ versus $l-a$ for the pressures $p=0.1, 0.3, 0.6$ and $1$
are pictured in figure \ref{fig2} by the solid/dashed curves for
the quantum/classical version of the model.
Quantum curves are approaching classical ones at higher temperatures
quite slowly; as follows from (\ref{ht}), the difference is of
order $1/\sqrt{T}$.
It is seen that NTE anomaly is absent for quantum 1D Tonks gas as well.
In other words, the quantization of classical model does not induce
NTE for 1D pure hard rods.

\renewcommand{\theequation}{4.\arabic{equation}}
\setcounter{equation}{0}

\section{Square well model} \label{Sec4}
The square well (SW) model is defined by the interaction potential
(\ref{general}) with \cite{Santos16}
\begin{equation} \label{SW}
\varphi(x) = -\varepsilon , \qquad a < \vert x\vert < a'
\end{equation}
where the well depth $\varepsilon>0$.
This soft-core potential is discontinuous at both $\vert x\vert = a$
and $\vert x\vert = a'$.

The Laplace transform (\ref{Om}) becomes \cite{Montero19}
\begin{equation} \label{SqW}
\widehat\Omega(s) = \frac{{\rm e}^{-a' s}+{\rm e}^{\beta \varepsilon}
({\rm e}^{-a s}-{\rm e}^{-a' s})}{s}
\end{equation}
and the reciprocal density (\ref{recdensity}) reads as
\begin{equation} \label{SqWl}
l(T,p) = a' + \frac{T}{p} - \frac{a'-a}{1-{\rm e}^{-(a'-a)p/T}
+ {\rm e}^{[-\varepsilon-(a'-a)p]/T}}.
\end{equation}

The low-temperature dependence of EoS (\ref{SqWl}) is simple:
\begin{equation} \label{SqWlT}
l = a + \frac{T}{p} + O({\rm e}^{-(a'-a)p/T}). 
\end{equation} 
Thus for $T=0$ the equidistant lattice spacing equals $a$, meaning
any positive pressure pushes the particles to the hard core.

\begin{figure}[t]
\begin{center}
\includegraphics[clip,width=0.84\textwidth]{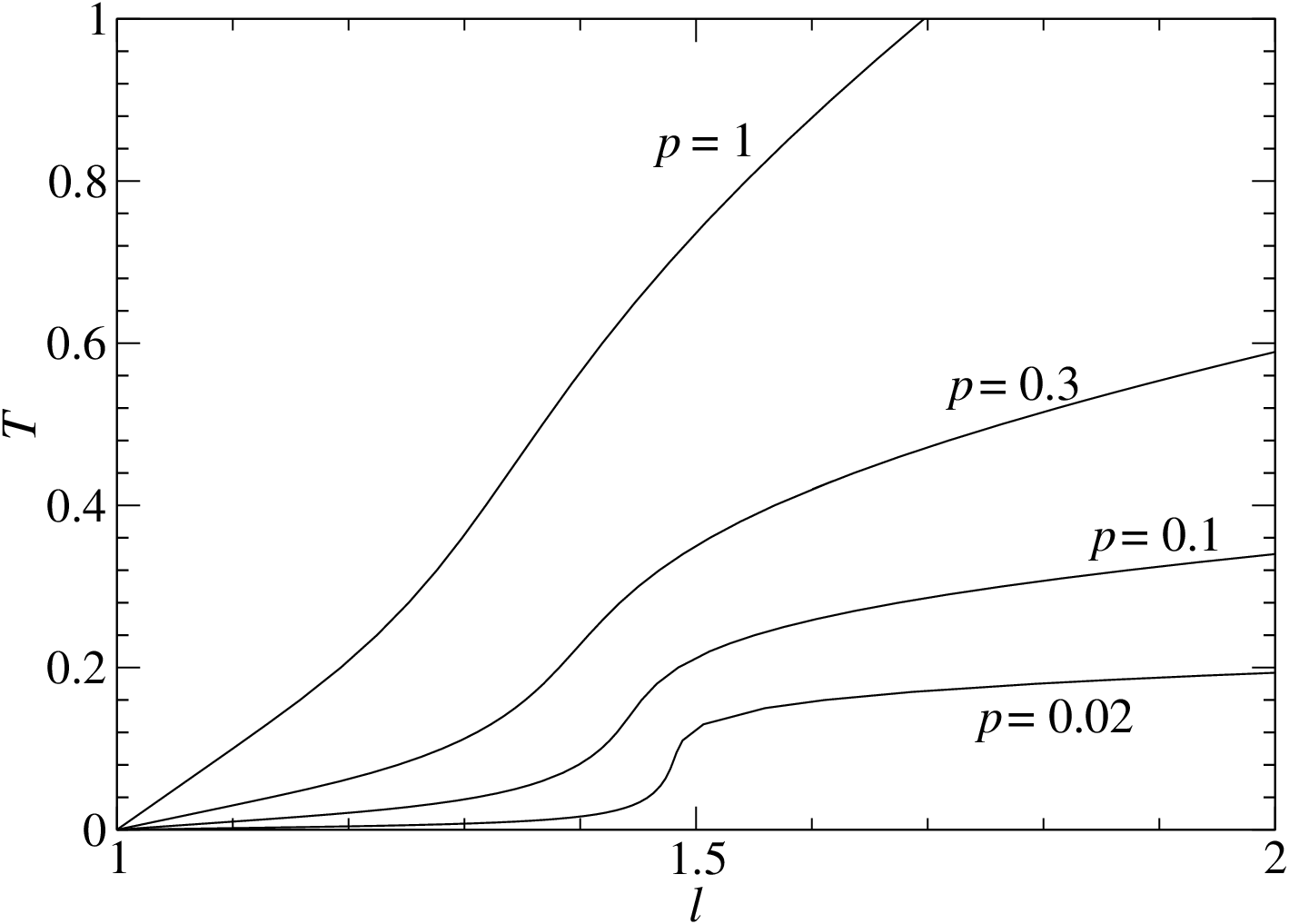}
\caption{SW model (\ref{SW}) with parameters $a=1$, $a'=2$ and
$\varepsilon=1$.
The plot shows the dependence of $T$ versus $l$ for four values of the pressure
$p=1,0.3,0.1$ and $0.02$.}
\label{fig3}
\end{center}
\end{figure}

Let us choose the length parameters as follows $a=1$, $a'=2$ and
the energy scale $\varepsilon=1$, as shown by the dotted curve
in figure \ref{fig1}.
The plots of $T$ versus $l$, given by (\ref{SqWl}), are shown in
figure \ref{fig3} for four values of the pressure $p=1,0.3,0.1$ and $0.02$.
Since $l$ grows monotonously with $T$ the anomalous NTE is absent for
this model.

\begin{figure}[t]
\begin{center}
\includegraphics[clip,width=0.84\textwidth]{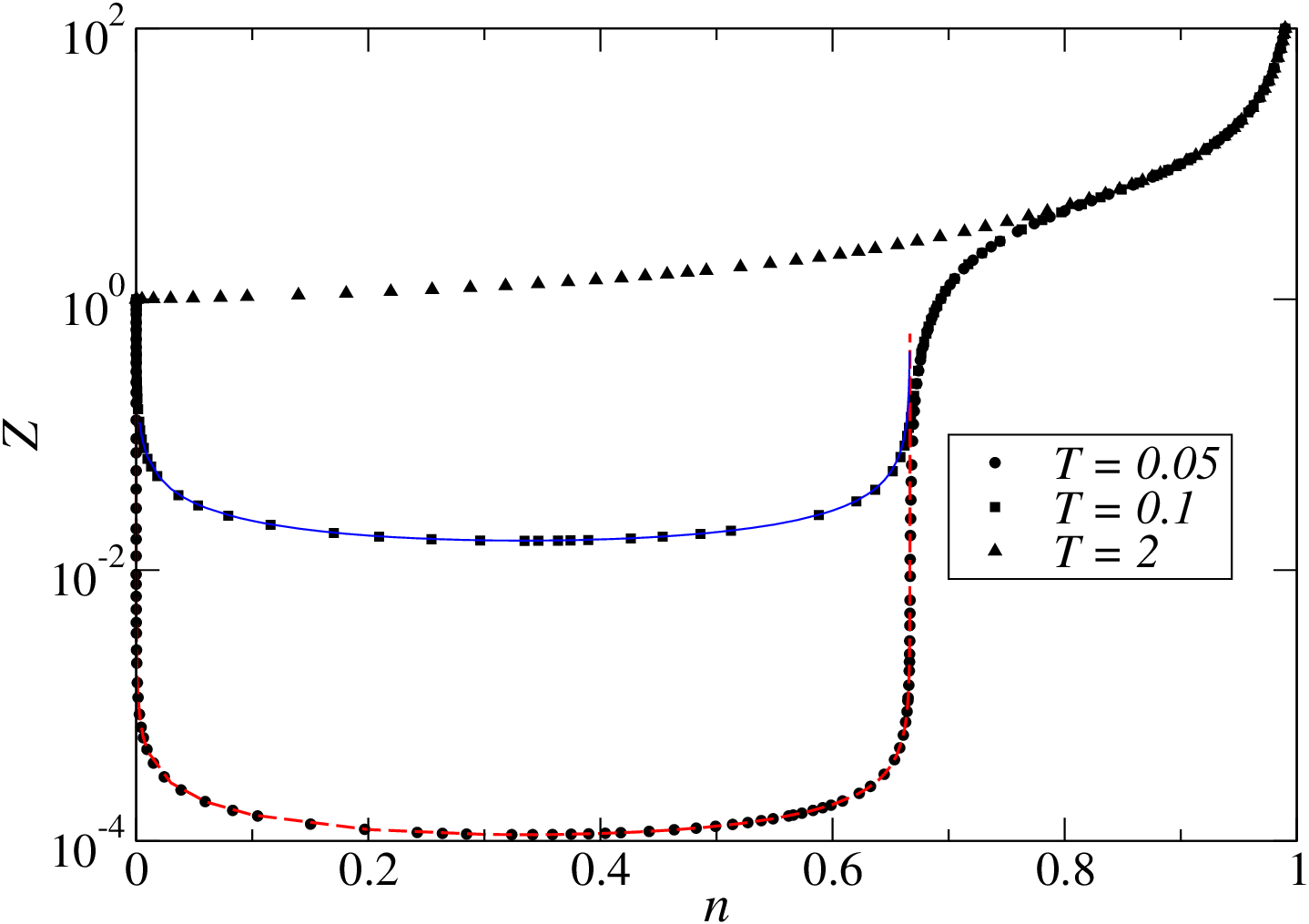}
\caption{SW model (\ref{SW}) with parameters $a=1$, $a'=2$ and $\varepsilon=1$.
The logarithmic plot of $Z$ versus $n$ is shown for three isotherms
$T=2$ (full triangles), $T=0.1$ [full squares with solid-line prediction
of (\ref{Znfit})] and $T=0.05$ [full circles with dashed-line prediction
of (\ref{Znfit})].
See the text for a detailed description of the plots.}
\label{fig4}
\end{center}
\end{figure}

The compressibility factor $Z$ is plotted as a function of the particle
density $n$ in figure \ref{fig4} for three isotherms:
$T=2$ (full triangles), $T=0.1$ (full squares) and $T=0.05$ (full circles).
The plot $Z(n)$ increases monotonously for the relatively large value of $T=2$.
For lower values of $T=0.05$ and $0.1$, $Z(n)$ exhibits a steep descent in
the region of very small $n$ followed by a wide plateau inside which
the pressure is very weak.
The steep climb to values of order of unity takes place at around
$n\approx 2/3$.
$Z(n)$ exhibits the leading singularity of the expansion (\ref{Znan})
when approaching $n\to 1^-$.

To derive a characteristic equation for the plateau region in the low-$T$
regime, let us introduce the small variable
\begin{equation}
\lambda\equiv \exp\left( -\beta\varepsilon\right)
\end{equation}  
and make an analytic analysis of the EoS (\ref{SqWl}) in the regime
$\lambda<\beta p (a'-a)<1$.
Equation (\ref{SqWl}) can be rewritten and treated as follows
\begin{eqnarray} 
Z & = & a'\beta p + 1 - \frac{\beta p(a'-a)}{1-{\rm e}^{-\beta p(a'-a)}}
\frac{1}{1+ \lambda \frac{1}{{\rm e}^{\beta p(a'-a)}-1}} \nonumber \\
& \approx & a'\beta p + 1 - \frac{1}{1-\frac{1}{2} \beta p (a'-a)}
\frac{1}{1+ \frac{\lambda}{\beta p(a'-a)}} , \label{SqWlnew}
\end{eqnarray}
where the Taylor expansion of the exponentials in the last term on the rhs
in powers of the small parameter $\beta p(a'-a)$ was used. 
Expanding once more the rhs of (\ref{SqWlnew}) in small parameters
$\beta p(a'-a)$ and $\lambda/[\beta p(a'-a)]$, and substituting
$\beta p = Z n$, we end up with
\begin{equation} \label{Znfit}
Z \approx \sqrt{\frac{\lambda}{n(a'-a)\left[1-\frac{a+a'}{2}n\right]}} .
\end{equation}
The validity of the formula $Z<1$ is equivalent to
the inequality
\begin{equation}
\frac{1}{2} \left( a'^2-a^2 \right) n^2 - (a'-a) n + \lambda < 0
\end{equation}
which leads to the condition for the density interval
\begin{equation}
n(a'-a) \in \left( \lambda, \frac{2(a'-a)}{a'+a} - \lambda \right) .
\end{equation}  
For our specific choice of $a=1$ and $a'=2$, the density border
$2(a'-a)/(a'+a)=2/3$ coincides with the point at which the isotherms
$T=0.1$ and $T=0.05$ climb, as shown in figure \ref{fig4}.
As is seen in the same figure, the prediction of formula (\ref{Znfit})
for temperatures $T=0.05$ (dashed curve) and $T=0.1$ (solid curve) is
very accurate within the density interval $n\in (0,2/3)$.

The (repulsive) square shoulder potential is derived from the (attractive)
SW potential by substituting $\varepsilon\to -\varepsilon$
$(\varepsilon>0)$ in equation (\ref{SW}).
Its exact EoS coincides with (\ref{SqWl}) under the same substitution
$\varepsilon\to -\varepsilon$.
Despite the formal similarity between the two potentials, their physics
are fundamentally different.
In a range of the pressures and low temperatures, the square shoulder
potential exhibits the NTE anomaly \cite{Travenec25}.
The plots of the compressibility factor $Z$ versus $n$ also show
distinct behaviors \cite{Travenec25}.

\renewcommand{\theequation}{5.\arabic{equation}}
\setcounter{equation}{0}

\section{Polynomial potentials with just one minimum} \label{Sec5}
Let us now study purely attractive soft-core potentials $\varphi(x)\le 0$
that possess a polynomial form.
The studied polynomials are of orders 2 and 3.

\subsection{Quadratic model} \label{Sec5.1}
Based on a standard 1D model of harmonically coupled oscillators, we consider
a 1D fluid with the quadratic interaction potential \cite{Percus87}
\begin{equation} \label{Q}
\varphi(x) = \varepsilon\frac{(\vert x\vert-a_m)^2}{(a'-a_m)^2}-\varepsilon ,
\qquad a< \vert x\vert < a' .
\end{equation}
The potential $\varphi(x)$ has a minimum at $x=a_m$ with
the value $\varphi(a_m)=-\varepsilon$.
At the potential border $x=a'$, while the soft-core potential is continuous, 
$\varphi(a')=0$, its first derivative $\varphi'(x)$ is not continuous.
Note that if one chooses $a_m=(a+a')/2$ the potential vanishes
at $x\to a^+$, $\varphi(x\to a^+)=0$.

The Laplace transform (\ref{Om}) of the Boltzmann factor with
the potential (\ref{Q}) takes the form
\begin{eqnarray} 
\hat\Omega(s) & = & \frac{{\rm e}^{-a' s}}{s} +
\frac{\sqrt{\pi}}{2\sqrt{\beta \varepsilon}}
(a'-a_m)\exp\left[\beta\varepsilon -a_ms +
\frac{(a'-a_m)^2 s^2}{4\beta \varepsilon}\right]
\nonumber\\ & & \times
\left\{ {\rm erf}\left[\frac{2\beta \varepsilon+(a'-a_m)s}{
2\sqrt{\beta \varepsilon}}\right]
-{\rm erf}\left[\frac{2(a_m-a)\beta \varepsilon-(a'-a_m)^2s}{
2(a'-a_m)\sqrt{\beta \varepsilon}} \right] \right\} ,
\nonumber\\ & & \label{QOm}
\end{eqnarray}
where
\begin{equation} \label{error}
{\rm erf}(z) = \frac{2}{\sqrt{\pi}} \int_0^z {\rm d}t\, {\rm e}^{-t^2}
\end{equation}
is the error function \cite{Gradshteyn}.
The rather complicated formula for the reciprocal density
(\ref{recdensity}) is not presented here.

\begin{figure}[t]
\begin{center}
\includegraphics[clip,width=0.84\textwidth]{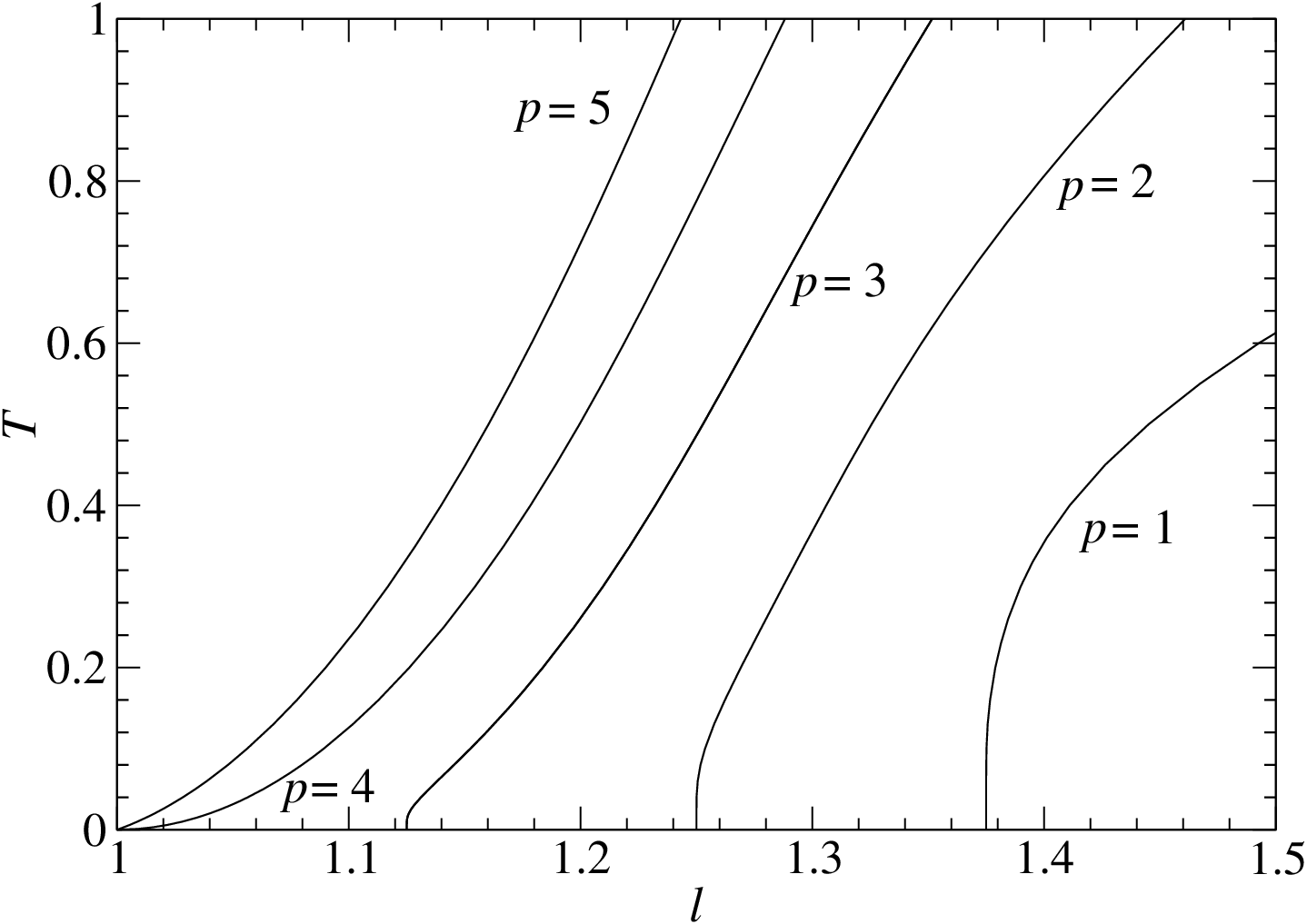}
\caption{Quadratic model with one minimum (\ref{Q}).
Parameters are chosen as follows $a=1$, $a_m=3/2$, $a'=2$, $\varepsilon=1$
and the incompressibility pressure $p_i=4$.
The dependence of $T$ versus $l$ is plotted for five values
of the pressure $p=5,4,3,2$ and $1$.}
\label{fig5}
\end{center}
\end{figure}

The ground-state spacing $l(T=0,p)$ for pressures $0<p<p_i$ can be deduced
from (\ref{derener}),
\begin{equation}
l(T=0,p) = x^*(p) = a_m - \frac{p(a'-a_m)^2}{2\varepsilon} , \qquad
\mbox{for $0<p<p_i$.}
\end{equation}
The incompressibility pressure, given by the equality $l(T=0,p_i)=a$,
reads as
\begin{equation} \label{ppi}
p_i = \frac{2\varepsilon (a_m-a)}{(a'-a_m)^2} .
\end{equation}  
The ground-state spacing remains equal to $a$ for pressures
larger than $p_i$,
\begin{equation}
l(T=0,p) = a , \qquad \mbox{for $p\ge p_i$.}
\end{equation}

The leading $T$-correction of the mean spacing $l(T,p)$ minus the
ground-state spacing $l(T=0,p)$ is given by
\begin{equation}\label{lqT}
l(T,p) - l(0,p) = \left\{
\begin{array}{ll}  
\displaystyle{\frac{T}{p-p_i} + O(T^2)} & \mbox{if $\displaystyle{p>p_i}$,} \cr
\displaystyle{(a'-a_m)\sqrt{\frac{T}{\pi\varepsilon}}+\cdots\phantom{aaaa}} &
\mbox{if $\displaystyle{p=p_i}$,} \cr
\cdots & \mbox{if $\displaystyle{p<p_i}$.}
\end{array} \right.
\end{equation}  
The dots represent exponentially small terms of type $\exp(-c/T)$ with $c>0$.
The linear $T$-correction for $p>p_i$ coincides with that derived from
(\ref{lpiTgreater}) and the square-root $\sqrt{T}$-correction for $p=p_i$
coincides with that derived from (\ref{lpiT}).
Since $\varphi''(x)$ does not depend on $x$, the derivative with respect to $p$
on the rhs of (\ref{lpT}) vanishes, signalizing the absence of the
term linear in $T$ (and also higher powers of $T$) for $p<p_i$.  

Choosing the parameters $a=1$, $a_m=3/2$, $a'=2$ and $\varepsilon=1$
in the definition of the quadratic potential (\ref{Q}),
we find that the value of the potential at the minimum point $\varphi(3/2)=-1$,
as shown by the dashed-dotted curve in figure \ref{fig1}. 
Formula (\ref{ppi}) tells us that the incompressibility pressure $p_i=4$.
The dependence of $T$ versus $l$ is plotted for five values
of the pressure $p=5,4,3,2$ and $1$ in figure \ref{fig5}.
It is evident that NTE is absent for this potential.

The isotherms of $Z$ versus $n$ are similar to those for the square well
model in figure \ref{fig4}.

\subsection{Cubic models} \label{Sec5.2}
By fixing the parameters $a=1$ and $a'=2$, we introduce the cubic potential
of the form
\begin{equation} \label{Cg}
\varphi(x) =\alpha_1(x-2)+\alpha_2(x-2)^2+\alpha_3(x-2)^3 
\end{equation}
ensuring the continuity of $\varphi(2)=0$ at $x=2$.
Here, $\alpha_1$, $\alpha_2$ and $\alpha_3$ are real coefficient.
Since
\begin{equation} \label{thirdder}
\varphi'''(x)= 6 \alpha_3 \qquad \mbox{for any $x\in [1,2]$,}
\end{equation}
the sign of the coefficient $\alpha_3$ determines the sign of the third
derivative of $\varphi(x)$.
We also require that $\varphi(1)=0$, leading to the constraint
\begin{equation} \label{eqq1}
\alpha_2 = \alpha_1 + \alpha_3 .
\end{equation}
The condition for the minimum point $a_m$ of the soft-core potential reads as
\begin{equation} \label{am}
\frac{\partial \varphi(x)}{\partial x} \Big\vert_{x=a_m}
= \alpha_1 + 2 \alpha_2 (a_m-2) + 3 \alpha_3 (a_m-2)^2 = 0 .   
\end{equation}
The minimum point is expected to be within the interval $1< a_m < 2$.
It is useful to introduce new parameter $t$ via
\begin{equation} \label{tam}
t \equiv 2 - a_m , \qquad t\in (0,1) , 
\end{equation}
which represents the deviation of the minimum point $a_m$ from the edge point 2.
In terms of $t$ equation (\ref{am}) can be rewritten as
\begin{equation} \label{eqq2}
\alpha_1 - 2 \alpha_2 t + 3 \alpha_3 t^2 = 0 .   
\end{equation}
The final requirement is that $\varphi(a_m)=-1$ which is equivalent to
the relation
\begin{equation} \label{eqq3}
-\alpha_1 t + \alpha_2 t^2 - \alpha_3 t^3 = - 1 .
\end{equation}

The three equations (\ref{eqq1}), (\ref{eqq2}) and (\ref{eqq3}) can be
solved to express the coefficients $\alpha_1$, $\alpha_2$ and $\alpha_3$
in terms of $t$:
\begin{equation}
\alpha_1 = \frac{2-3t}{t(1-t)^2} , \quad
\alpha_2 = \frac{1-3t^2}{t^2(1-t)^2} , \quad
\alpha_3 = \frac{1-2t}{t^2(1-t)^2} .
\end{equation}
Consequently,
\begin{eqnarray}
t\in \left( 0,\frac{1}{2} \right): & & \alpha_1,\alpha_2,\alpha_3>0 ,
\nonumber\\
t\in \left( \frac{1}{2},\frac{1}{\sqrt{3}} \right): & & \alpha_1,\alpha_2>0 ,
\quad \alpha_3<0 ,
\nonumber\\
t\in \left( \frac{1}{\sqrt{3}},\frac{2}{3} \right): & & \alpha_1>0 ,
\quad \alpha_2,\alpha_3<0 , \nonumber\\
t\in \left( \frac{2}{3},1 \right): & & \alpha_1\alpha_2,\alpha_3<0 .
\end{eqnarray}
The most significant change of sign is that of the coefficient $\alpha_3$
which determines the sign of the third derivative of $\varphi(x)$
inside the whole region $x\in [1,2]$, see equation (\ref{thirdder}).
The positive values of $\alpha_3$ correspond to the region of $t\in [0,1/2]$,
i.e., according to the relation (\ref{tam}), to the location of
the minimum point on the right side from the interval center $3/2$,
$a_m\in (3/2,2)$. 
Such geometry of the interaction potential, depicted in figure \ref{fig1} as
``Cubic sharp to the right'' of its minimum, forces particles to diminish their
distance due to thermal fluctuations which indicates presence of NTE.
On the other hand, the negative values of $\alpha_3$ correspond to the region
of $t\in [1/2,1]$, meaning the minimum point is located to the left side
of the interval center $3/2$, $a_m\in (1,3/2)$. 
Such geometry of the interaction potential, depicted in figure \ref{fig1} as
``Cubic sharp to the left'' of its minimum, forces particles to move further
apart from each other due to thermal fluctuations, indicating absence of NTE.

Applying of the condition for the minimum point of the $f$-function
(\ref{derener}) to the cubic potential (\ref{Cg}) yields a second-degree
equation for $x^*$
\begin{equation}
p + \alpha_1 + 2 \alpha_2 (x^*-2) + 3 \alpha_3 (x^*-2)^2 = 0 .  
\end{equation}
The physically acceptable solution is
\begin{eqnarray}
x^*(p,t) & = & \frac{1}{3(1-2t)} \Big\{ 5 - 12 t + 3 t^2 \nonumber \\
& & + \sqrt{1-6t+3(5-p)t^2-6(3-2p)t^3+(9-15p)t^4+6pt^5} \Big\} .
\nonumber \\ & & \label{xstar} 
\end{eqnarray}  
The plus sign ahead of the square root was chosen to ensure
that for $p=0$ (when $x^*=a_m$) it holds that $x^*=2-t$, in agreement
with definition (\ref{tam}).

\subsubsection{Cubic model with $\varphi'''(x)<0$} \label{4.2.1}
Let us first consider negative coefficients $\alpha_3$,
say the case $t=3/5$ when
\begin{equation} \label{C}
\varphi(x) =\frac{25}{36}\left[3(x-2)-2(x-2)^2-5(x-2)^3\right].
\end{equation}
The minimum point of this potential occurs at $a_m=7/5$ and
$\varphi(a_m)=-1$ as it should be, see the dashed curve in figure \ref{fig1}. 
This cubic potential is sharp to the left of its minimum.
It does not satisfy Kuzkin's condition (\ref{Kuzkin}) and therefore
the absence of NTE anomaly is anticipated.
A similar form and the absence of NTE is a common feature of many potentials,
including the Lennard-Jones potential.

\begin{figure}[t]
\centering
\setbox1=\hbox{\includegraphics[clip,height=7.5cm]{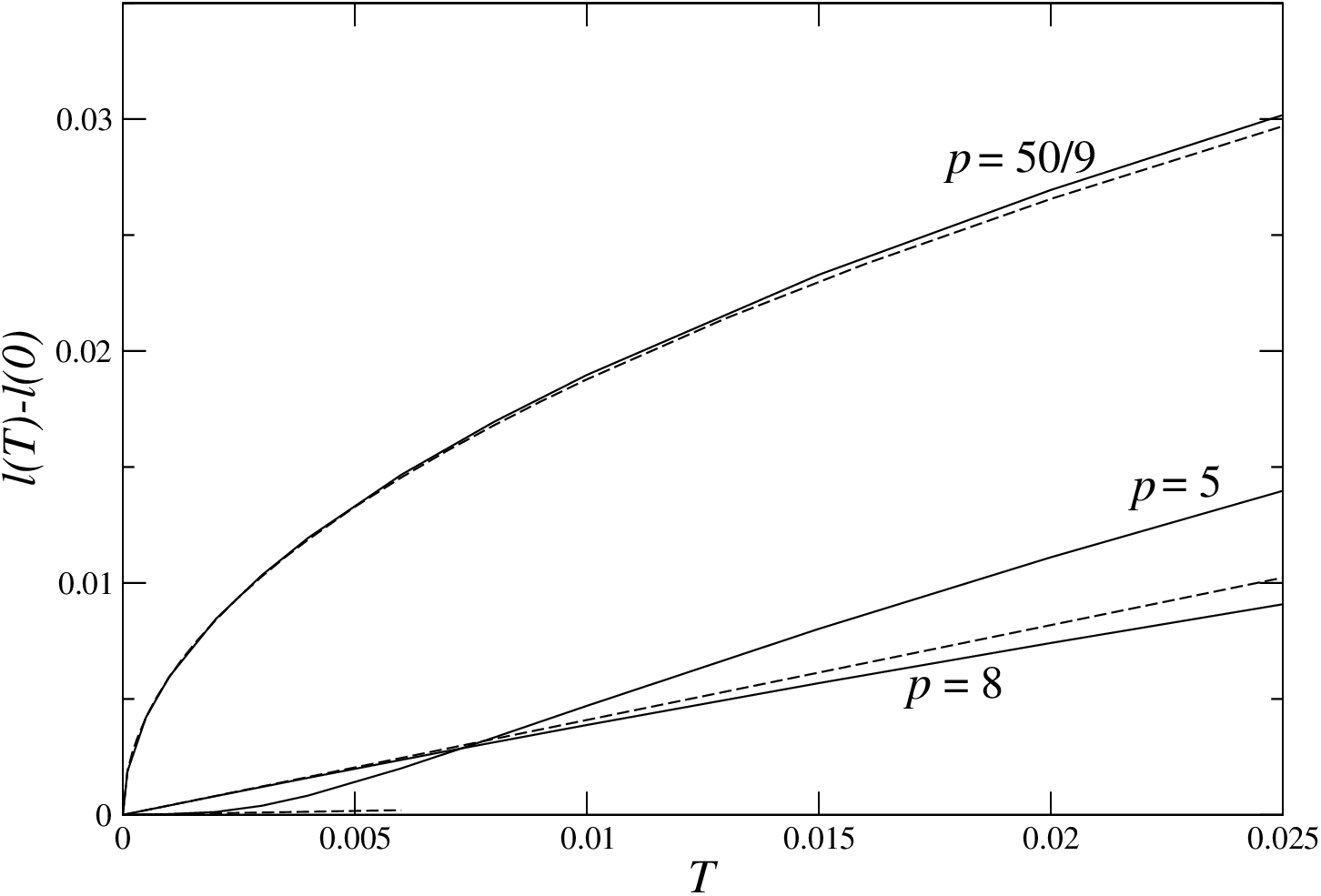}}
\includegraphics[clip,height=7.5cm]{figure6a.eps}\lapbox[5cm]{-9.9cm}
{\raisebox{4.7cm}{\includegraphics[clip,height=2.7cm]{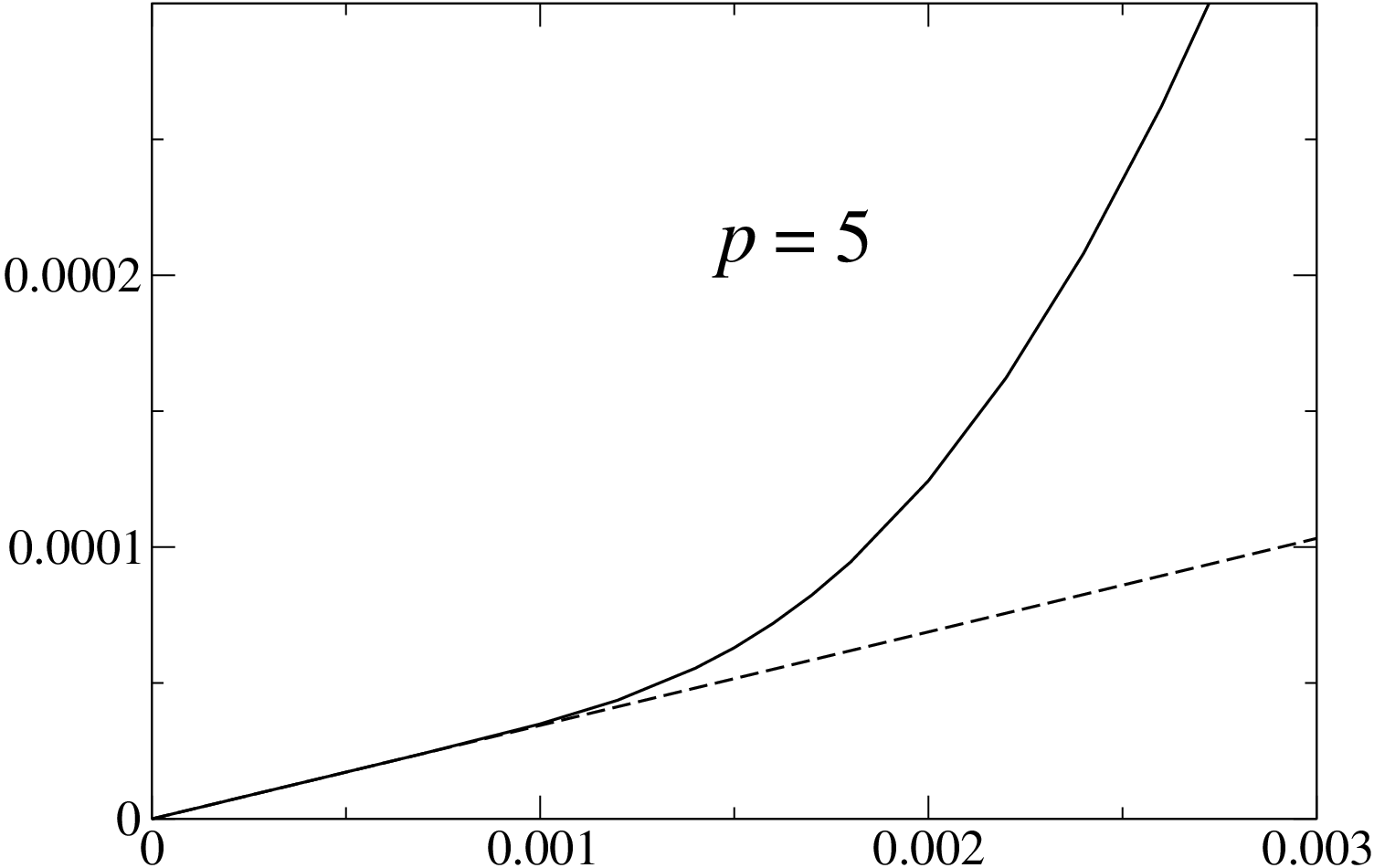}}}
\caption{Results for the cubic potential (\ref{C}) with negative
third derivative.
The dependence of $l(T)-l(0)$ versus $T$ is pictured for three values of
the pressure $p=8$, $p=p_i=50/9=5.555...$ and $p=5$ by solid curves.
Dashed lines describe the plots given by the leading terms in (\ref{lCT}).
The inset magnifies the very-low-temperature region for $p=5$.}
\label{fig6}
\end{figure}

Integrals in the exact EoS (\ref{recdensity}) can only be treated numerically
for the present potential.
Nevertheless the ground state and leading terms of
the low-$T$ expansion of $l(T,p)$ can be found analytically.
The substitution of $t=3/5$ into equation (\ref{xstar}) yields  
\begin{equation}
x^*(p) = \frac{1}{75} \left( 140 - \sqrt{5}\sqrt{245+108p} \right) .
\end{equation}
This $x^*(p)$ corresponds to the ground-state spacing between
particles $l(T=0,p)$ and varies continuously from the minimum point of
the potential $a_m=7/5$ when $p\to 0$ up to the smallest possible value
$x=1$ at the incompressibility pressure $p_i=50/9$.
In summary:
\begin{equation}
l(0,p) = \left\{
\begin{array}{lll}
1 & & \mbox{if $p>p_i$,} \cr
\displaystyle{\frac{1}{75} \left( 140 - \sqrt{5}\sqrt{245+108p} \right)}
& & \mbox{if $0<p<p_i$.}
\end{array} \right.
\end{equation}  
The coefficients of the leading low-$T$ expansion of the difference
$l(T,p)-l(0,p)$ can be obtained analytically from the previous formulas
(\ref{lpT}), (\ref{lpiT}) and (\ref{lpiTgreater}):
\begin{equation}\label{lCT}
l(T,p) - l(0,p) = \left\{
\begin{array}{lll}  
\displaystyle{\frac{1}{p-p_i} T + O(T^2)} & &
\mbox{if $\displaystyle{p>p_i}$,} \cr
\displaystyle{\frac{6}{5}\frac{1}{\sqrt{13\pi}} \sqrt{T}
+ o\left( \sqrt{T}\right)} & & \mbox{if $\displaystyle{p=p_i}$,} \cr
\displaystyle{\frac{27}{245+108p} T + O\left( T^2\right)} & &
\mbox{if $\displaystyle{0<p<p_i}$.}
\end{array} \right.
\end{equation}
Their check obtained by a numerical treatment of formula (\ref{recdensity})
is presented in figure \ref{fig6} for three
values of the pressure $p=8$, $p=p_i=50/9=5.555...$ and $p=5$
(dashed lines) against the full values of the difference $l(T,p)-l(0,p)$
(solid lines).
Since the leading $T$-term has an extremely small prefactor for $0<p<p_i$,  
the inset magnifies the very-low-temperature region for $p=5$.

\begin{figure}[t]
\begin{center}
\includegraphics[clip,width=0.84\textwidth]{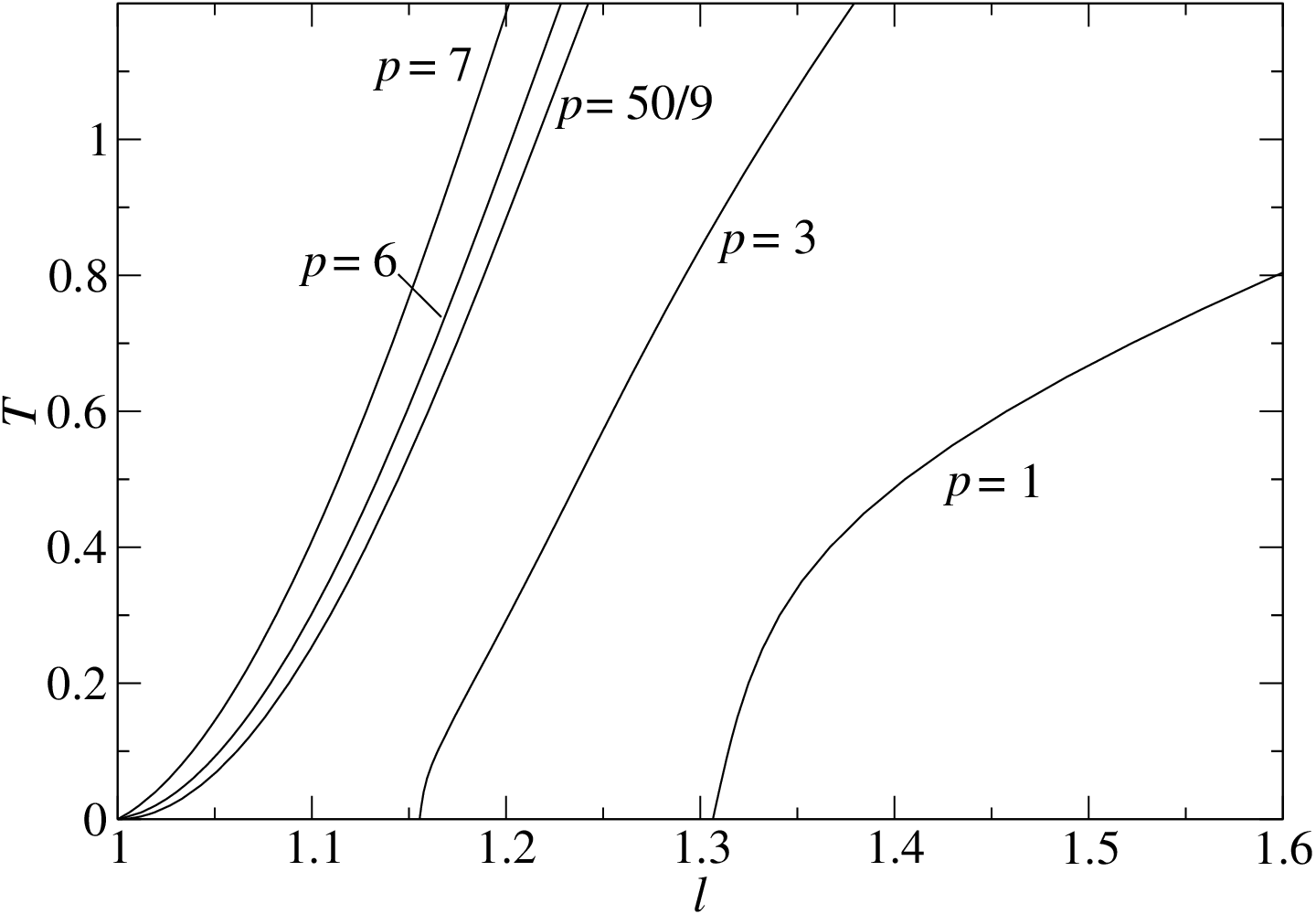}
\caption{Cubic model (\ref{C}) with negative third derivative.
The plot shows the relationship between $T$ and $l$ for five values of
pressure $p=7,6,50/9,3$ and $1$.
The value $50/9$ corresponds to the compressibility pressure $p_i$.}
\label{fig7}
\end{center}
\end{figure}

The dependence of $T$ versus $l$ is plotted for five values of
the pressure $p=7,6,50/9,3$ and $1$ in figure \ref{fig7}.
The plots do not exhibit the anomalous NTE behavior as was expected.  

The isotherm curves of the dependences $Z(n)$ are similar to those
of the SW model in figure \ref{fig4}.

\subsubsection{Cubic model with $\varphi'''(x)>0$} \label{Sec5.2.2}
Now, let's consider positive coefficients $\alpha_3$,
say the case $t=1/3$ when
\begin{equation} \label{C2}
\varphi(x) =\frac{27}{4}\left[x-2+2(x-2)^2+(x-2)^3\right].
\end{equation}
The minimum point of this potential occurs at $a_m=5/3$, see
the solid curve in figure \ref{fig1}. 
This cubic potential is sharp to the right of its minimum
and satisfies Kuzkin's condition (\ref{Kuzkin}), making it
a candidate for NTE anomaly.

\begin{figure}[t]
\begin{center}
\includegraphics[clip,width=0.84\textwidth]{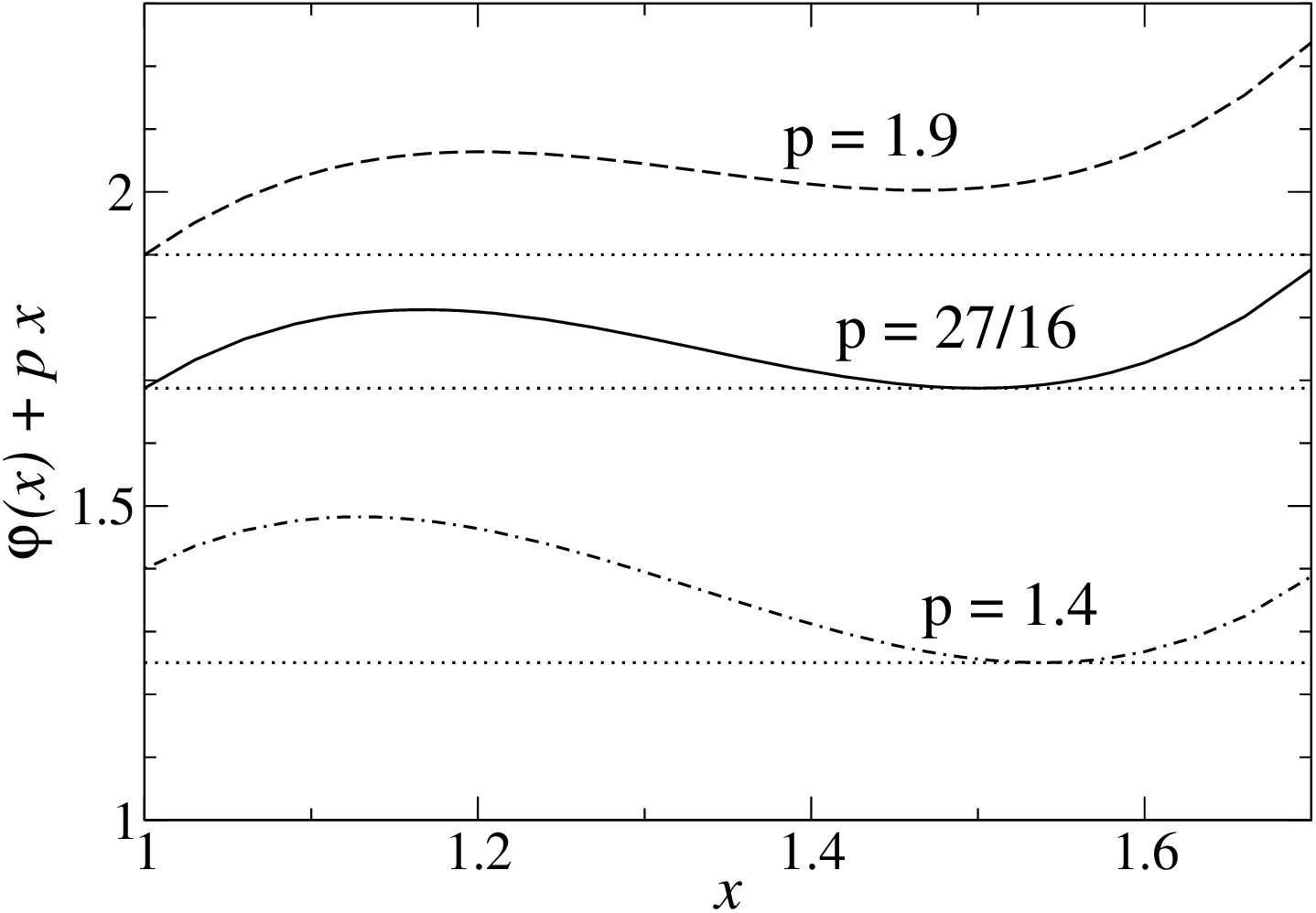}
\caption{The cubic model (\ref{C2}) with positive third derivative.
The plot shows the function $f(x)=\varphi(x)+px$ for three
pressure values: $p=1.4$ (dash-dotted curve), $p=27/16=1.6875\ldots$
(solid curve) and $p=1.9$ (dashed curve).
Dotted lines indicate the minimum value of $f(x)$ for a given $p$
on the interval $x\in [1,2]$.}
\label{fig8}
\end{center}
\end{figure}

Substituting $t=1/3$ into equation (\ref{xstar}) results in  
\begin{equation} \label{aa}
x^*(p) = \frac{1}{9} \left( 12 + \sqrt{9 - 4 p} \right) .
\end{equation}
Similar to the previous case, this ground-state spacing between
particles $l(0,p)$ moves continuously from the minimum point of
the potential $a_m=5/3$ as $p\to 0$.
This continuous movement stops at specific ``jump''
pressure $p_j$ before $x^*(p)$ reaches its smallest possible value 1.
This scenario is displayed in figure \ref{fig8} where the plot of the
function $f(x)=\varphi(x)+px$ (which needs to be minimized with respect to $x$
in the interval $1\le x\le 2$) is shown for three values of pressure:
$p=1.4$ (dash-dotted curve), $p=27/16=1.6875\ldots$ (solid curve) and
$p=1.9$ (dashed curve).
When $p=1.4$ the only minimum point for $f(x)$ is given by (\ref{aa}).
When the pressure reaches its jump value $p_j=27/16$ the minimum of $f(x)$ is
doubly degenerate at two points $x^*(p_j)=3/2$ and $x=1$.
Increasing pressure beyond $p_j$, there is only one minimum of $f(x)$
at $x=1$.
Consequently, there is a jump between the minimums $x^*(p_j)=3/2$ and
$x=1$ at $p=p_j$.
The jump value of pressure $p_j=27/16$ is determined by the relation
\begin{equation} \label{jumprelation}
\varphi\left(x^*(p_j)\right) + p_j x^*(p_j) = \varphi(1) + p_j ,
\end{equation}
where
\begin{equation} \label{aaa}
x^*(p_j) = \frac{1}{9} \left( 12 + \sqrt{9 - 4 p_j} \right) .
\end{equation}

To derive the value of the jump pressure for an arbitrary value of
$t$, one inserts the expression (\ref{xstar}) for the general
$x^*(p,t)$ into equation (\ref{jumprelation}), resulting in
\begin{equation} \label{eqa}
p_j(t) = \frac{(2-3t)^2}{4(1-t)^2(1-2t)} .
\end{equation}
As a check we reproduce $p_j=27/16$ for the above case $t=1/3$. 
It is clear from figure \ref{fig8} that the jump scenario occurs
only if $f'(1)\ge 0$ at $p=p_j(t)$, i.e.
\begin{equation} \label{eqb}
p_j(t) \ge \frac{3t-1}{(1-t) t^2} .
\end{equation}  
By combining equations (\ref{eqa}) and (\ref{eqb}), we find that $t$ must be
from the interval $[0,1/2)$, meaning the jump exists for all cubic potentials
with positive third derivatives.

\begin{figure}[t]
\centering
\setbox1=\hbox{\includegraphics[clip,height=8.2cm]{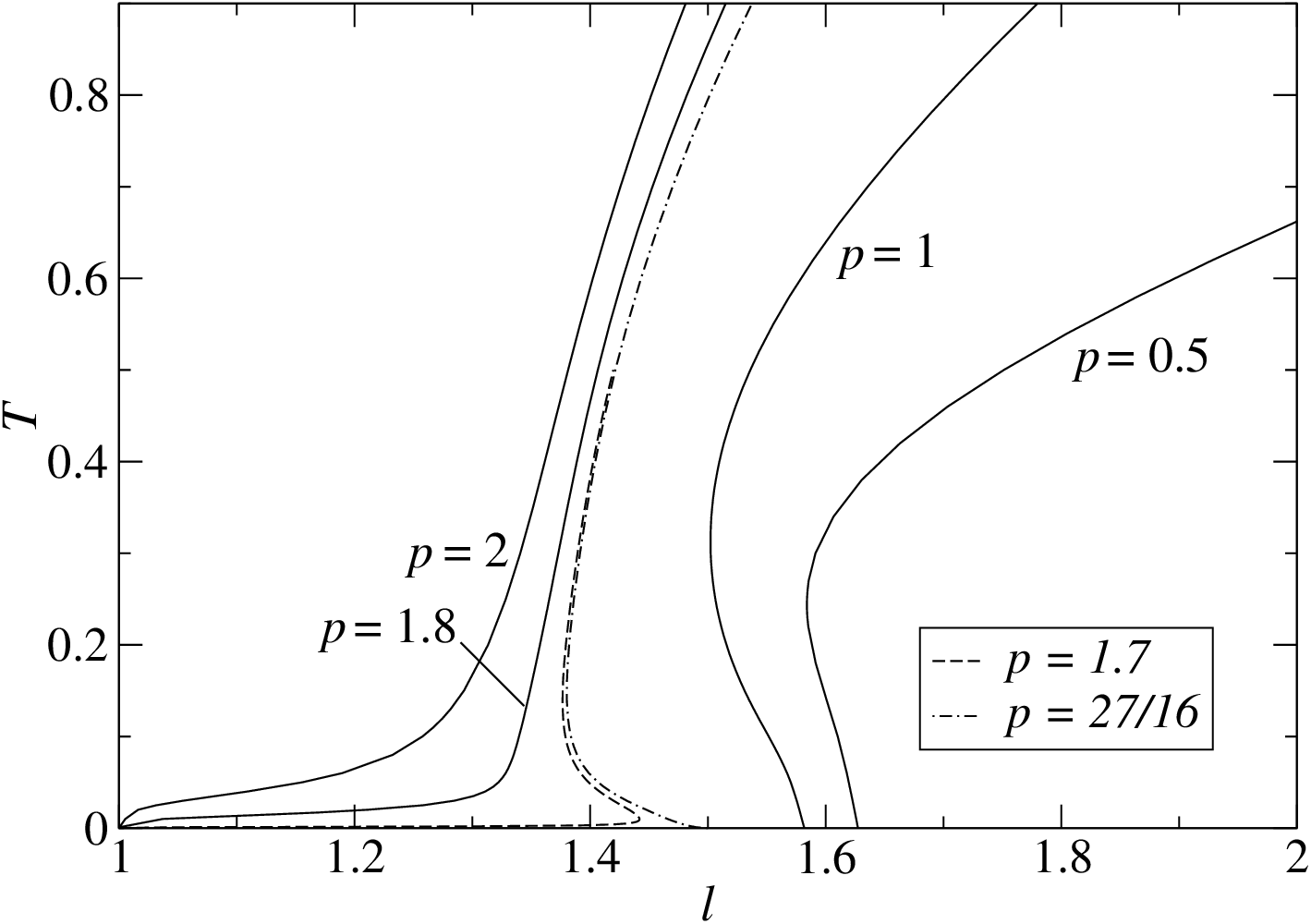}}
\includegraphics[clip,height=8.2cm]{figure9a.eps}\lapbox[5cm]{-10.3cm}
{\raisebox{5.5cm}{\includegraphics[clip,height=2.6cm]{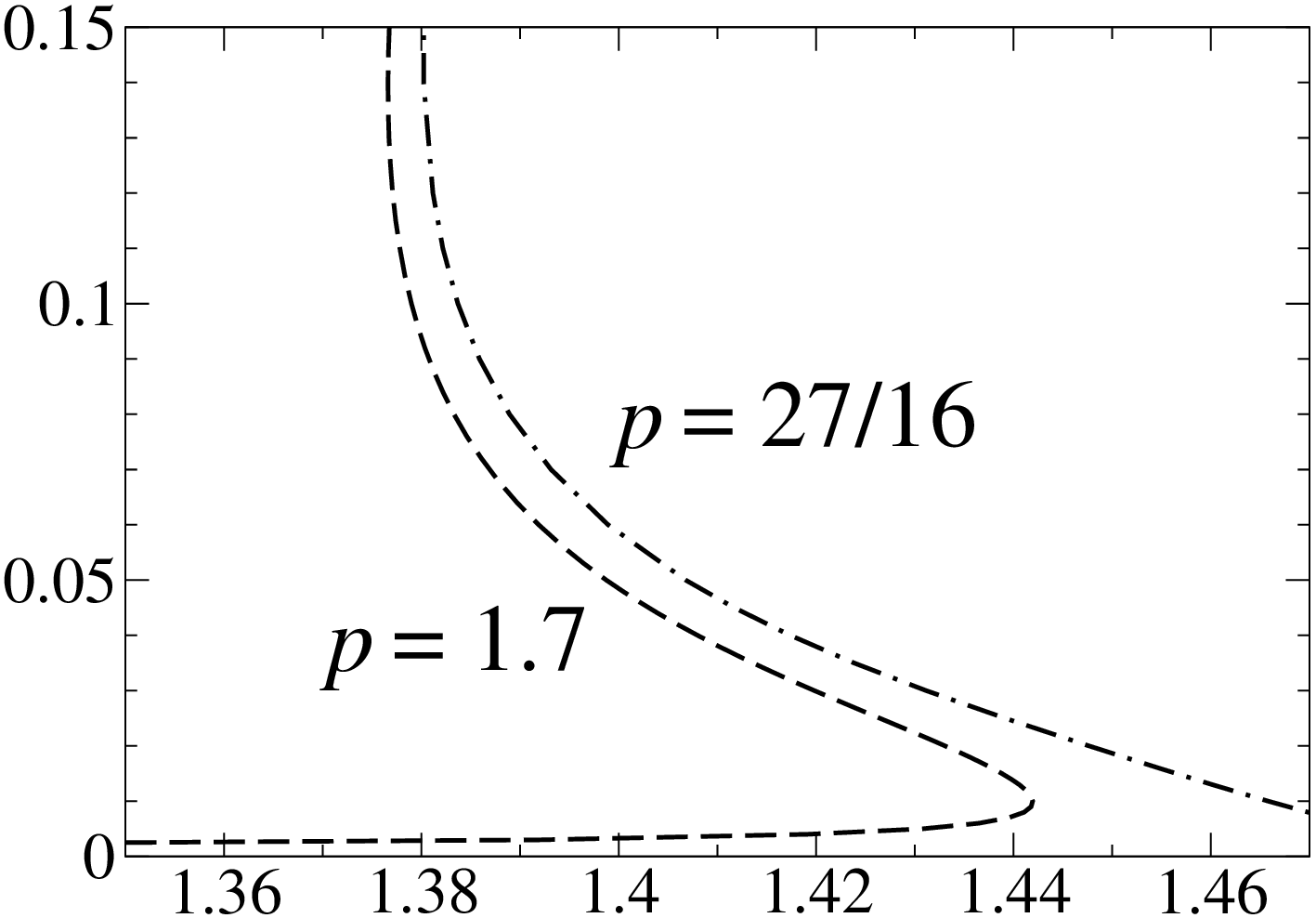}}}
\caption{Results for the cubic model (\ref{C2}) with positive third derivative.
The dependence of $T$ versus $l$ for six distinct values of the pressure
$p=2,1.8,1.7,27/16,1$ and $0.5$.
The inset magnifies the low-temperature region around $l\approx 1.4$
for two closely related pressure values $p_j=27/16=1.6875\ldots$
(dash-dotted curve) and $p=1.7$ (dashed curve).
Refer to the text for a detailed description of the plots.}
\label{fig9}
\end{figure}

The formulas that include the ground state plus the leading low-$T$
correction are written as 
\begin{equation} \label{lCasR}
l(T,p) = \left\{
\begin{array}{lll}  
\displaystyle{1+\frac{T}{p}+O(T^2)} & & \mbox{if $\displaystyle{p>27/16}$,} \cr
\displaystyle{\frac{1}{9}\left(12+\sqrt{9-4p}\right)-\frac{T}{9-4p}+O(T^2)} &
& \mbox{if $\displaystyle{0<p\le 27/16}$.}
\end{array} \right.
\end{equation}
For $0<p\le 27/16$, the prefactor $-1/(9-4p)$ to the linear-$T$ term
was derived from (\ref{lpT}) and it was also checked numerically.
It is important to note  that this prefactor is negative throughout
the interval $0<p\le 27/16$, providing a direct evidence of
the NTE phenomenon.  
For $p>27/16$, the prefactor $1/p$ to the linear-$T$ term was obtained
with a high precision numerically.

The plots of $T$ versus $l$ for six values of the pressure
$p=2,1.8,1.7,27/16,1$ and $0.5$ are presented in figure \ref{fig9}.
As is clear from equation (\ref{lCasR}), for $0<p\le p_j=27/16=1.6875\ldots$
the negative prefactor to the linear $T$-correction to $l(0)$
ensures the presence of NTE in the region of low temperatures,
starting from $T=0$.
As documented in the figure on the case $p=1.7$, NTE also remains in
an interval of $p>p_j$, up to approximately $p\approx 1.75$, in spite of
the positive prefactor $1/p$ to the to the linear $T$-correction to $l(0)$;
NTE exists in an interval of low {\em strictly positive} temperatures
excluding $T=0$.
The comparison of the curves for two close values of the pressure
$p=27/16=1.6875\ldots$ (dash-dotted curve) and $p=1.7$ (dashed curve)
in the inset of the figure indicates that NTE is related to the jump
in $l(0)$ at $p_j$ for $p\in [p_j,1.75]$.

\begin{figure}[t]
\begin{center}
\includegraphics[clip,width=0.84\textwidth]{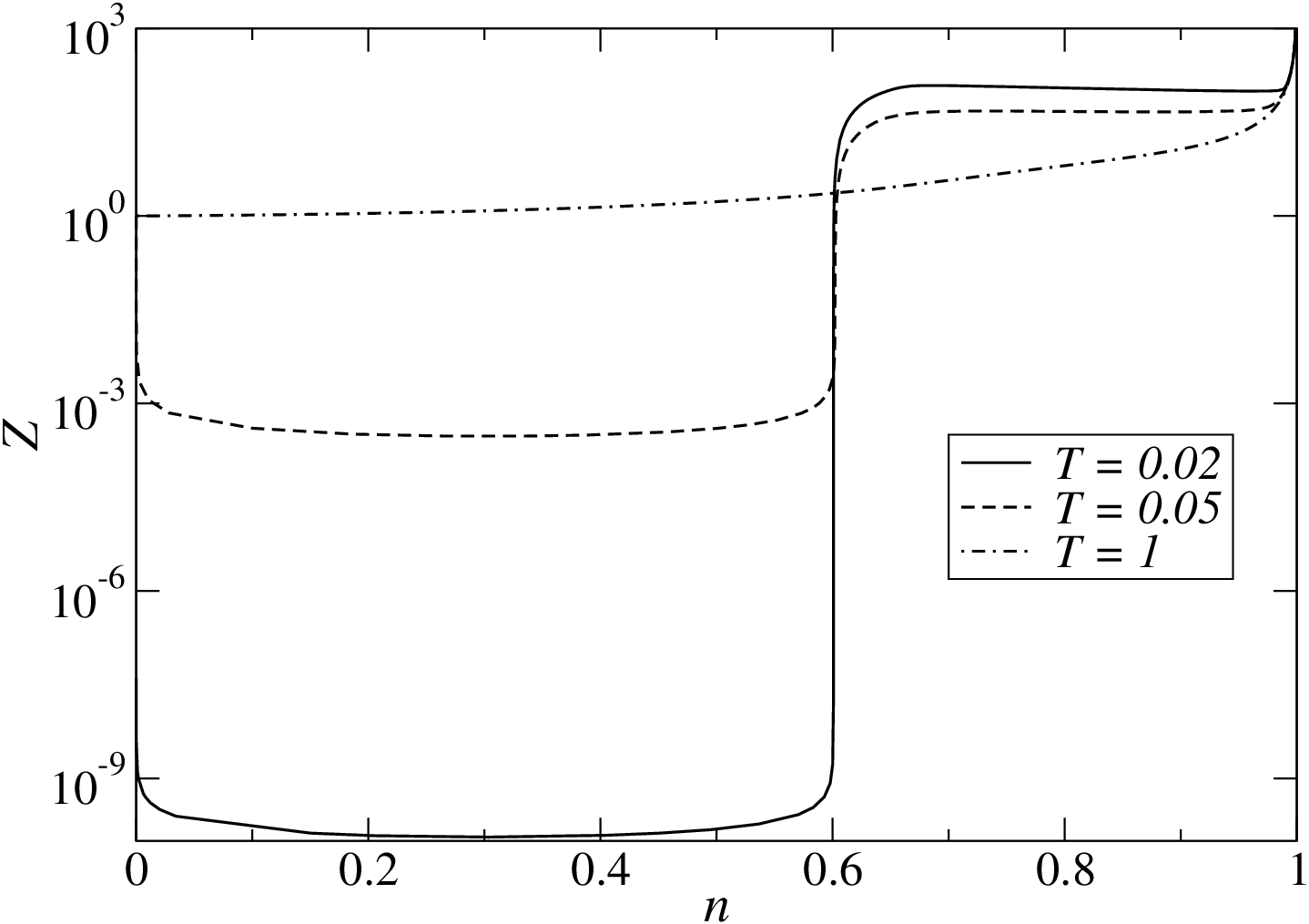}
\caption{Cubic model sharp to the right of its minimum (\ref{C2}).
The dependence of $Z$ versus $n$ is plotted for three fixed values of
the temperature $T=1, 0.05$ and $0.02$.}
\label{fig10}
\end{center}
\end{figure}

The dependence of $Z$ versus $n$ is plotted for three fixed values of
the temperature $T=1, 0.05$ and $0.02$ in figure \ref{fig10}.
For very low temperatures $0.02$ and $0.05$, the plot $Z(n)$ exhibits
two plateaus separated by a steep rise at $n\approx 1/a_m=0.6$ and
not very pronounced maximum and minimum around the higher plateau
before transitioning into the asymptotic behavior
$Z\approx 1/(1-n)$ in the limit $n\to 1^-$.

\renewcommand{\theequation}{6.\arabic{equation}}
\setcounter{equation}{0}

\section{Models with non-analytic potentials} \label{Sec6}

\subsection{Fusion of two linear segments} \label{Sec6.1}
The interaction potential of this model is defined by
(\ref{general}) with $\varphi(x)$ consisting of two linear segments
connected at the minimum $x=a_m$ with $\varphi(a_m)=-\varepsilon$
$(\varepsilon>0)$ and fulfilling the boundary conditions
$\varphi(a)=\varepsilon_a$ and $\varphi(a')=0$:
\begin{equation} \label{lrmin}
\varphi(x) = \left\{
\begin{array}{lll}
\displaystyle{-\frac{\varepsilon_a+\varepsilon}{a_m-a}\vert x\vert
+\frac{a_m\varepsilon_a+a\varepsilon}{a_m-a}} & &
\mbox{if $a<\vert x\vert<a_m$,} \cr
\displaystyle{\frac{\varepsilon}{a'-a_m}\vert x\vert
-\frac{a'\varepsilon}{a'-a_m}} & &
\mbox{if $a_m<\vert x\vert<a'$}
\end{array} \right.
\end{equation} 
and $\varepsilon_a>-\varepsilon$.
It is continuous for $\vert x\vert > a$, but its first derivative
with respect to $x$ is discontinuous, besides the border point
$\vert x\vert = a'$, also at the minimum point $\vert x\vert = a_m$;
denoting by $a_n$ the position of the non-analyticity point one has $a_n=a_m$.

The Laplace transform (\ref{Om}) of the Boltzmann factor of
the potential (\ref{lrmin}) and the reciprocal density (\ref{recdensity})
can be calculated analytically, but we do not present rather cumbersome
formulas here.
The incompressibility pressure $p_i=(\varepsilon+\varepsilon_a)/(a_m-a)$
is given by the slope of the first linear part of the interaction potential
(\ref{lrmin}).

In the ground-state, the particle spacing behaves as
\begin{equation} \label{lrT0}
l(T=0,p) = \left\{
\begin{array}{ll}
\displaystyle{a} & \mbox{if $\displaystyle{p>p_i}$,} \cr
\displaystyle{\frac{a+a_m}{2}} & \mbox{if $\displaystyle{p=p_i}$,} \cr
\displaystyle{a_m} & \mbox{if $p<p_i$.}
\end{array} \right.   
\end{equation}
This means that the particles form a close packed equidistant array with
the hard-core spacing $a$ for large pressures $p>p_i$. 
Below the transition pressure $0<p<p_i$, the spacing becomes equal to $a_m$,
i.e., the pressure is not able to kick a particle out of the potential minimum.
The transition value $(a+a_m)/2$ at $p=p_i$ corresponds to a mixture of
pair particle spacings $a$ and $a_m$ with the same density $1/2$.
The ground-state prescription (\ref{derener}) cannot be applied
to this model.

The leading temperature term is always linear in $T$,
$l(T,p)-l(0,p)\sim \alpha(p) T$ in the limit $T\to 0^+$, where the prefactor
$\alpha$ depends on the region of the pressure:
\begin{equation} \label{lr}
\alpha(p) = \left\{
\begin{array}{ll}
\displaystyle{\frac{1}{p-p_i}} & \mbox{if $\displaystyle{p>p_i}$,} \cr
\displaystyle{\frac{1}{2} \frac{(a_m-a)(a'-a_m)}{(a'-a_m)\varepsilon_a
+(a'-a)\varepsilon}} & \mbox{if $\displaystyle{p=p_i}$,} \cr
\displaystyle{\frac{(a'-a_m)(\varepsilon_a+\varepsilon)
-(a_m-a)[\varepsilon+2(a'-a_m)p] }{[\varepsilon_a+\varepsilon
-(a_m-a)p][\varepsilon+(a'-a_m)p]}} & \mbox{if $p<p_i$.}
\end{array} \right.   
\end{equation}
The prefactor is positive for $p>p_i$, diverging as $p$ approaches
the value $p_i$ from above.
It is finite and positive at $p=p_i$.
The next correction terms are exponentially small, of type $\exp(-c/T)$
with $c>0$, except for the transition pressure $p_i$ when $l(T,p_i)$ exhibits
Taylor's expansion in integer powers of $T$.

\begin{figure}[t]
\begin{center}
\includegraphics[clip,width=0.84\textwidth]{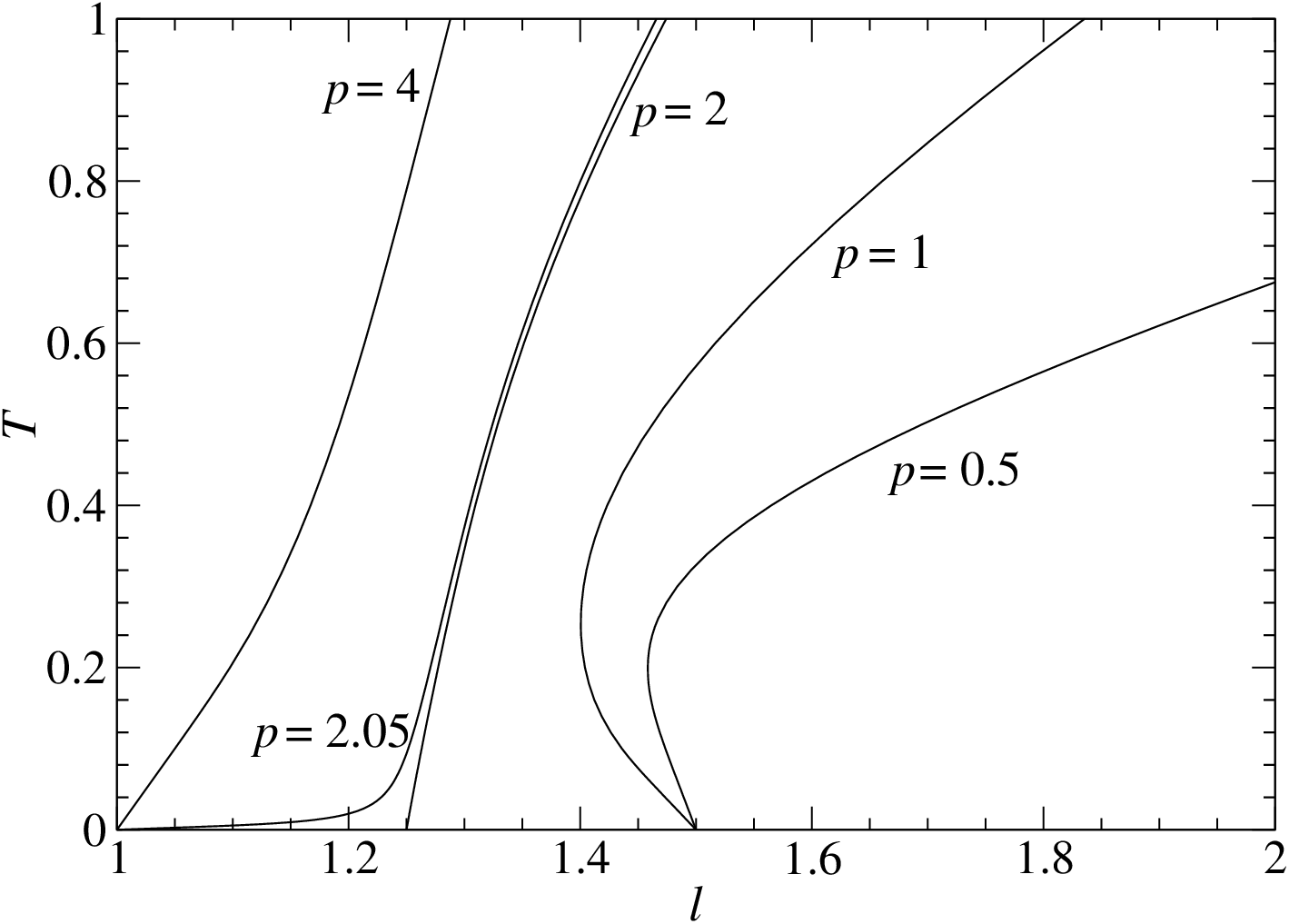}
\caption{The potential resulting from the  fusion of two
linear segments (\ref{lrmin}).
Parameters are chosen as follows $a=1$, $a_m=3/2$, $a'=2$, $\varepsilon=1$,
$\varepsilon_a=0$ and the incompressible pressure $p_i=2$.
The dependence of $T$ versus $l$ is plotted for five fixed values
of the pressure $p=4,2.05,2,1$ and $0.5$.
Refer to the text for a detailed description of the plots.}
\label{fig11}
\end{center}
\end{figure}

For the chosen parameters $a=1$, $a_m=3/2$, $a'=2$, $\varepsilon_a=0$
and $\varepsilon=1$, the transition pressure is $p_i=2$.
The low-$T$ expansion formula (\ref{lr}) now takes the form
\begin{equation} \label{lrrr}
l(T,p) = \left\{
\begin{array}{ll}
\displaystyle{1 + \frac{T}{p-2} + \cdots} &
\mbox{if $\displaystyle{p>2}$,} \cr
\displaystyle{\frac{5}{4} + \frac{T}{8}+O(T^2)} &
\mbox{if $\displaystyle{p=2}$,} \cr
\displaystyle{\frac{3}{2} - \frac{p}{2\left[ 1-(p/2)^2\right]} T + \cdots} &
\mbox{if $\displaystyle{0<p<2}$.}
\end{array} \right.   
\end{equation}
Note that the prefactor to $T$ is negative if $0<p<2$ which is connected
with existence of NTE in this interval of pressures.
The plots of $T$ versus $l$ are shown for five fixed values of the pressure
$p=4,2.05,2,1$ and $0.5$ in figure \ref{fig11}.
We can see NTE for $0<p<2$ as well as a jump of $l(0,p)\to 5/4$ at
$p=p_i=2$, see (\ref{lrrr}).

At low temperatures, the plot of $Z(n)$ shows two plateaus separated by
a steep rise at $n\approx1/a_m=2/3$, similar to figure \ref{fig10}.
This behavior is typical for models that exhibit a jump in $l(0,p)$.

\subsection{Fusion of linear and quadratic models with non-analyticity point
at $x=a_n<a_m$} \label{Sec6.2}
Let's choose specific values for $a=1$, $a_m=3/2$, $a'=2$ and
the non-analyticity point connecting linear and quadratic parts of
the interaction potential to the left of $a_m$, namely at
$a_n=5/4=1.25<a_m$.
Along with the conditions $\varphi(1)=\varphi(2)=0$, we have
\begin{equation} \label{phiLQ}
\varphi(x) = \left\{
\begin{array}{lll}  
3-3x & & \mbox{if $1<x<5/4$,}  \cr
\displaystyle{4\left(x-\frac{3}{2}\right)^2-1} & & \mbox{if $5/4<x<2$.}
\end{array} \right.
\end{equation} 
Thus $\varphi(5/4)=-3/4$ is continuous, but its derivative
$\varphi(x)'\vert_{x=5/4}$ is discontinuous.
The Laplace transform (\ref{Om}) of the Boltzmann factor of
the potential (\ref{phiLQ}) reads as
\begin{eqnarray} \label{LQOm}
\hat\Omega(s) & = & \frac{{\rm e}^{-2 s}}{s} +
\frac{{\rm e}^{(3\beta-5s)/4}-{\rm e}^{-s}}{3\beta-s}+
\frac{{\rm e}^{\beta-3s/2+s^2/(16\beta)}\sqrt{\pi}}{4\sqrt{\beta}}
\nonumber\\ & & \times
\left\{ {\rm erf}\left[\frac{4\beta+s}{4\sqrt{\beta}}\right]
-{\rm erf}\left[\frac{s-2\beta}{4\sqrt{\beta}} \right]\right\}
\end{eqnarray}
which yields a lengthy explicit expression for $l(T,p)$
which is not presented here.

\begin{figure}[t]
\begin{center}
\includegraphics[clip,width=0.84\textwidth]{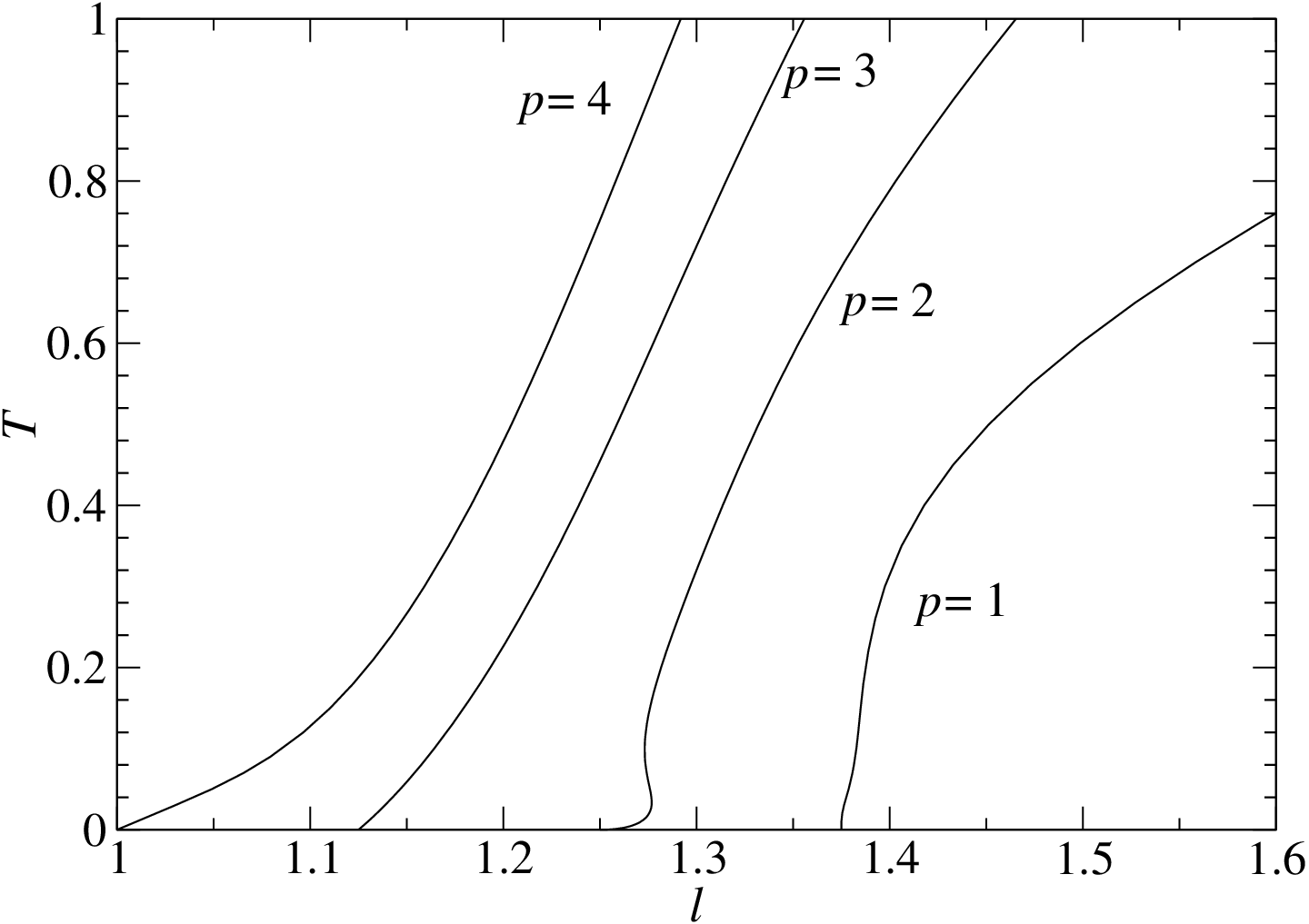}
\caption{Fusion of linear and quadratic models (\ref{phiLQ}).
The dependence of $T$ versus $l$ is plotted for four fixed values of
the pressure $p=4,3,2$ and $1$.
The transition pressure $p_i=3$ and NTE is seen at $p_n=2$.}
\label{fig12}
\end{center}
\end{figure}

After some algebra the low-$T$ expansion of $l(T,p)$ is found to be
\begin{equation}\label{lLQT}
l(p,T) = \left\{
\begin{array}{ll}  
\displaystyle{1+\frac{T}{p-3}+ O(T^2)} & \mbox{if $\displaystyle{p>3}$,} \cr
\displaystyle{\frac{9}{8}+\frac{T}{2}-2T^2+O(T^3)} &
\mbox{if $\displaystyle{p=3}$,} \cr
\displaystyle{\frac{5}{4}+\frac{2(p-5/2)}{(p-2)(p-3)}T+ O(T^2)} &
\mbox{if $\displaystyle{2<p<3}$,} \cr
\displaystyle{\frac{5}{4}-\frac{4}{\sqrt{\pi}}T^{3/2}+O(T^2)} &
\mbox{if $\displaystyle{p=2}$,} \cr
\displaystyle{\frac{3}{2}-\frac{p}{8}+\cdots} & \mbox{if
$\displaystyle{p<2}$.}
\end{array} \right.
\end{equation}  
The dots mean an exponentially small term.
There are two special pressures: $p_i=3$ matches the slope of the linear part
of $\varphi(x)$ and $p_n=2$ is given by the non-analyticity point $x=5/4$
which limits the validity of equation (\ref{derener}) to
$l(0,p_n)=3/2-p_n/8=5/4$.
The value $l(0)=9/8=(1+5/4)/2$ is the arithmetic mean at $p=p_i$.
Note that the subleading terms tend to diverge when the pressure approaches
special values, namely $p\to p_n^+$ and $p\to p_i$ from both sides.

The plots of $T(l)$ are shown in figure \ref{fig12} for pressures
$p=1$, $p_n=2$, $p_i=3$ and $p=4$. 
The NTE anomaly appears for medium pressures, for example, $p=2$.

\subsection{Fusion of quadratic and linear models with non-analyticity point
at $x=a_n>a_m$} \label{Sec6.3}
We choose special values $a=1$, $a_m=3/2$, $a'=2$ again and the non-analyticity
point connecting linear and quadratic part of interaction at $a_n=7/4>a_m$.
Together with the choice $\varphi(1)=\varphi(2)=0$, the soft-core potential
is given by
\begin{equation} \label{phiLQr}
\varphi(x) = \left\{
\begin{array}{lll}
\displaystyle{4\left(x-\frac{3}{2}\right)^2-1} & & \mbox{if $1<x<7/4$,} \cr
3x-6 & & \mbox{if $1<x<7/4$.}
\end{array} \right.
\end{equation} 
Thus $\varphi(x)$ is continuous at $x=7/4$, but its first derivative
is discontinuous at this point.

\begin{figure}[t]
\centering
\setbox1=\hbox{\includegraphics[clip,height=8.5cm]{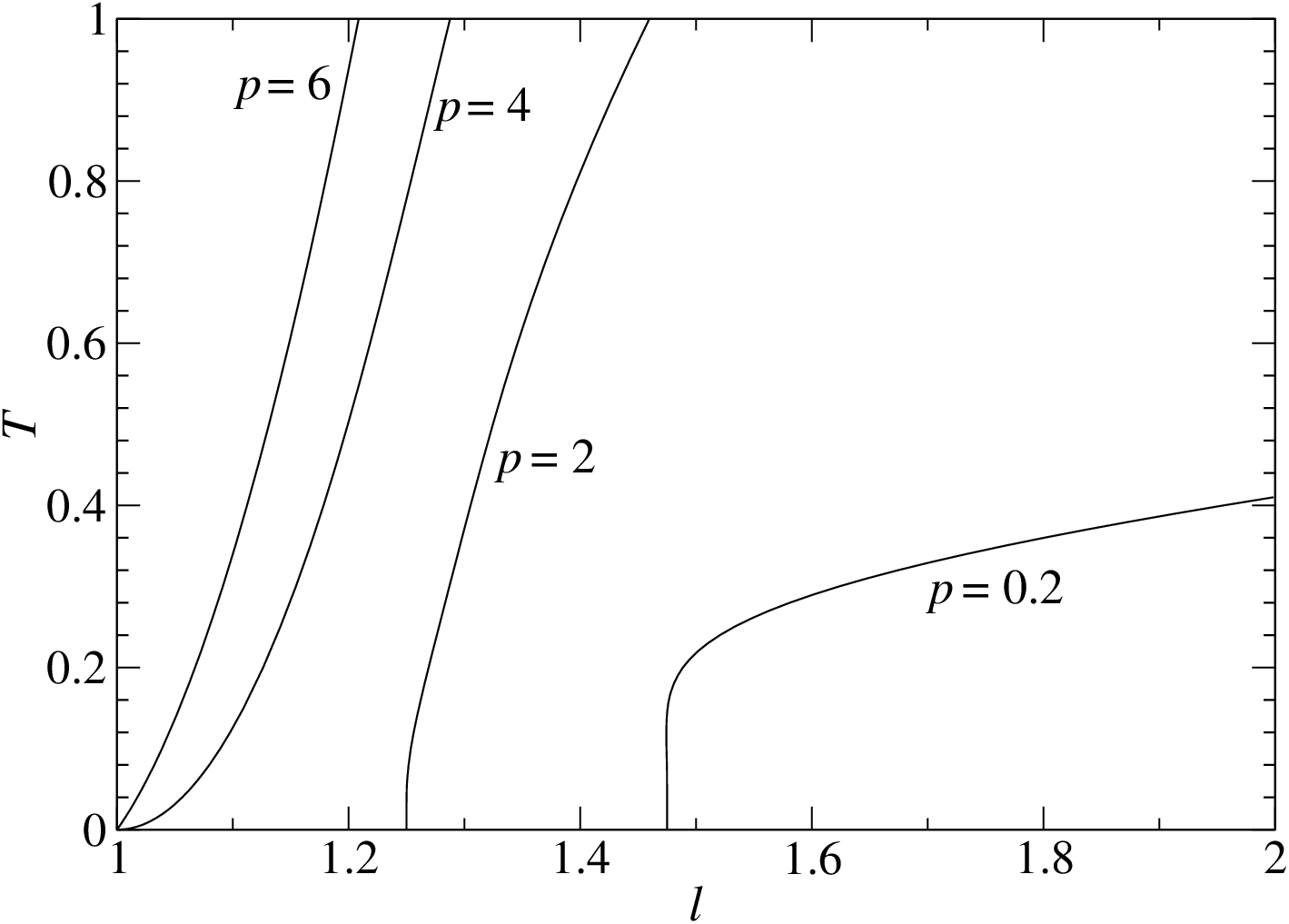}}
\includegraphics[clip,height=8.5cm]{figure13a.eps}\lapbox[5cm]{-6cm}
{\raisebox{4.1cm}{\includegraphics[clip,height=3.9cm]{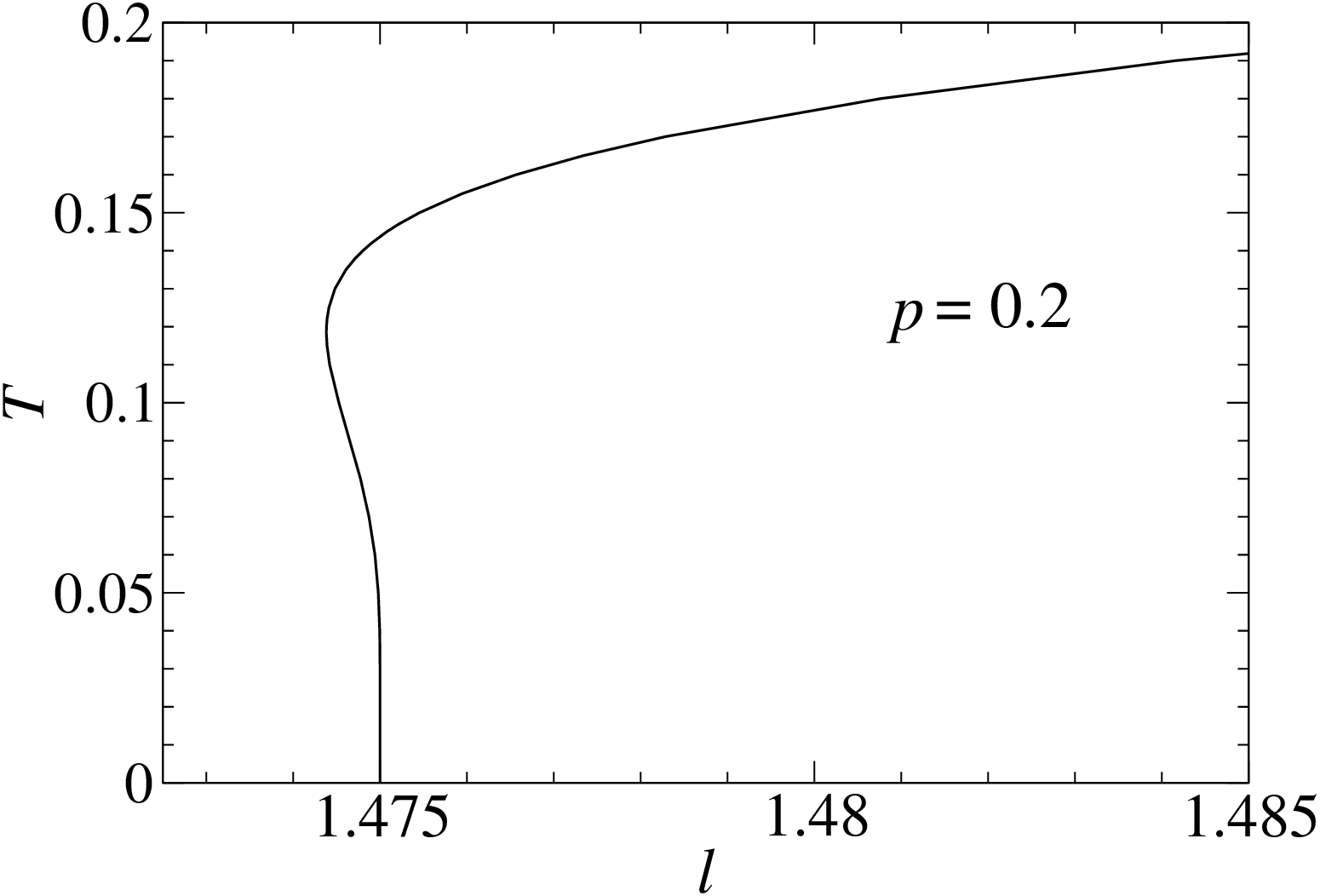}}}
\caption{Numerical results for the fusion model (\ref{phiLQr}).
The dependence of $T$ versus $l$ is shown for four different values
of the pressure $p=6,4,2$ and $0.2$.
The inset magnifies the low-temperature region around $l(T=0,p=0.2)$
characterized by the presence of NTE.}
\label{fig13}
\end{figure}

The Laplace transform (\ref{Om}) of the Boltzmann factor of
the potential (\ref{phiLQr}) reads as
\begin{eqnarray} \label{LQrOm}
\hat\Omega(s) & = & \frac{{\rm e}^{-2 s}}{s} +
\frac{{\rm e}^{-2 s}\left[{\rm e}^{(3\beta+s)/4}-1\right]}{3\beta+s}+
\frac{{\rm e}^{\beta-3s/2+s^2/(16\beta)}\sqrt{\pi}}{4\sqrt{\beta}}
\nonumber\\ & & \times
\left\{ {\rm erf}\left[\frac{4\beta-s}{4\sqrt{\beta}}\right]
+{\rm erf}\left[\frac{2\beta+s}{4\sqrt{\beta}}
\right]\right\} .
\end{eqnarray}
This equation provides the expression for $l(T,p)$ which is not included here.
The ground state and low-$T$ expansion of $l(T,p)$ can be written as
\begin{equation}\label{lLQrT}
l(T,p) = \left\{
\begin{array}{lll}  
\displaystyle{1+\frac{T}{p-4}-\frac{16\ T^2}{(p-4)^3}+ O(T^3)} & &
\mbox{if $\displaystyle{p>4}$,} \cr
\displaystyle{1+\frac{1}{2\sqrt{\pi}}\sqrt{T}+\cdots} & &
\mbox{if $\displaystyle{p=4}$,} \cr
\displaystyle{\frac{3}{2}-\frac{p}{8}+\cdots} & &
\mbox{if $\displaystyle{p<4}$.}
\end{array} \right.
\end{equation}  
Here, the dots represent exponentially small terms.
The incompressibility pressure $p_i=4$ can be derived from (\ref{derener})
at the hard core $l(0,p_i)=3/2-p_i/8=1$.

In figure \ref{fig13} we plot $T(l)$ for pressures $p=0.2$, $2$, $p_i=4$
and $p=6$. 
We still observe NTE for medium pressures, such as $p=0.2$
but it is barely visible as shown in the inset.

\renewcommand{\theequation}{7.\arabic{equation}}
\setcounter{equation}{0}

\section{Conclusion} \label{Sec7}
Previous observations of the NTE anomaly in 2D and 3D systems of particles
with anisotropic and isotropic pairwise interactions were limited
to experiments and numerical simulations.
Our previous work \cite{Travenec25} on 1D hard rods with various types
of soft repulsive nearest-neighbor interaction potentials allowed us to
treat the NTE phenomenon based on exact thermodynamic results.
The potentials have two competing length scales: the diameter of
hard cores $a$ (dominant at high temperatures and pressures)
and the finite range $a'$ of the soft component constrained by
$a\le a'\le 2a$ (dominant at low temperatures and pressures). 
While we did not identify a precise mathematical criterion for the potential's
characteristics that imply NTE, there are conditions that facilitate
the emergence of NTE.
One of these conditions is the jump in chain spacing $a'\to a$ of
the equidistant ground state at certain pressures.
Another condition is the non-analyticity of the soft potential
$\varphi(x)$ within the interval $x\in (a,a')$.
It is important to note that Kuzkin's necessary and sufficient condition
for the presence of NTE (\ref{Kuzkin}) does not apply to these
repulsive particle systems.

The present paper continues the discussion of NTE phenomenon in other
exactly solvable 1D fluids.

First question studied in section \ref{Sec3} is whether the quantization
of a classical fluid (which is free of NTE) can lead to NTE.  
To answer this question we use classical pure hard rods of diameter $a$
which do not exhibit NTE and possess simple ground state:
for an arbitrary pressure $p>0$ the particle spacing equals to $a$, see
section \ref{Sec3.1}.
As is shown in section \ref{Sec3.2}, due to the uncertainty principle
the quantum version of pure hard rods posseses a more complicated ground
state with the particle spacing dependent on pressure $p$,
see the relation (\ref{quantumgs}).
In spite of a shift of the ground-state spacing with respect to $a$
for finite $p>0$, the quantum curves of dependence $T$ versus $l-a$
for fixed $p$ (depicted in figure \ref{fig2} by solid lines) are going to
their classical counterparts (depicted in figure \ref{fig2} by dashed lines)
at infinite $T$ monotonously. 
This fact suggests an unimportant role of quantum mechanics in the generation
of NTE in more general fluid models.

Secondly, the classical hard rods with various types of soft nearest-neighbor
interactions that contain a basin of attraction with only one minimum
are investigated.
NTE occurs in two cases for such systems.
\begin{itemize}
\item
As shown in section \ref{Sec5.2} for cubic potentials, NTE
is present if Kuzkin's condition (\ref{Kuzkin}) applies.
It seems that in the derivation of Kuzkin's condition the presence of
an attractive basin for the potential is a necessary condition
from the beginning.
This also means that it is a sufficient condition but not a necessary one.
The NTE phenomenon is accompanied by a jump in the equidistant ground state in
spacing at a certain pressure $p_j$.
\item
As shown in section \ref{Sec6}, NTE occurs as soon as
the soft potential $\varphi(x)$ is non-analytic at a point within
the interval $x\in(a,a')$.
The position of the non-analyticity point $a_n$ with respect to the minimum
point of the potential $a_m$ is irrelevant from this perspective,
refer to section \ref{Sec6.1} for $a_n=a_m$, section \ref{Sec6.2} for $a_n<a_m$
and section \ref{Sec6.3} for $a_n>a_m$.
\end{itemize}

The compressibility factor $Z(\beta,n)$, introduced in section \ref{Sec2.3}, 
is a thermodynamic quantity that exhibits a nontrivial behavior for
our 1D fluids with nearest-neighbor interactions. 
In the case of purely repulsive interaction potentials \cite{Travenec25},
isotherms of $Z$ as a function of the particle density $n$ exhibit both
monotonous and non-monotonous plots.
In the present case of attractive interaction potentials, low-temperature
isotherms of $Z(n)$ exhibit a steep descent for very small $n$,
followed by a wide plateau up to $n\approx 2/3$ (for our choice of parameters)
within which the pressure is very weak, as seen in
figure \ref{fig4} for the SW model.
For the cubic model sharp to the right of its minimum, $Z(n)$ exhibits
a double-plateau structure, as shown in figure \ref{fig10}.

The present treatment can be extended to mixtures, such as
the polydisperse Tonks gas \cite{Santos16,Evans10}.

It is known that degenerate Bose versions of 1D fluids can be prepared
experimentally via a confinement in optical lattices
\cite{Kinoshita04,Haller10} and micro-traps \cite{Kruger10,Langen15}. 
The Bethe-ansatz solution of the 1D Tonks gas \cite{Wadati02,Samaj13}
can serve as a guide for an exact solution of a more general 1D quantum gas,
for example, with square shoulder or well nearest-neighbor interactions.
  
\begin{acknowledgements}
The support received from VEGA Grant No. 2/0089/24 is acknowledged.
\end{acknowledgements}

\section*{Declarations}

\begin{itemize}
\item Funding   Not applicable 
\item Conflict of interest/Competing interests   Not applicable
\item Ethics approval and consent to participate   Not applicable
\item Consent for publication   Not applicable   
\item Data availability   Data are available upon a request. 
\item Materials availability   Not applicable
\item Code availability   Not applicable 
\item Author contribution   Not applicable
\end{itemize}

\end{document}